\documentclass[default]{aastex7}
\usepackage{gensymb}
\usepackage{amsmath}
\usepackage{booktabs}
\usepackage{graphicx}
\usepackage{amssymb}
\hypersetup{linkcolor=magenta,citecolor=blue,filecolor=blue,urlcolor=blue}

\defcitealias{Hunt2024}{HR24}

\newcommand{\gaia}{\textit{Gaia}}
\newcommand{\vhel}{$v\rm_{\scriptscriptstyle helio}$}
\newcommand{\vlsr}{$v\rm_{\scriptscriptstyle LSR}$}
\newcommand{\kms}{$\rm km\,s^{-1}$}
\newcommand{\masyr}{$\rm mas\,yr^{-1}$}
\newcommand{\pmra}{$\mu_{\alpha^{*}}$}
\newcommand{\pmdec}{$\mu_{\delta}$}
\newcommand{\plx}{$\varpi$}
\newcommand{\g}{$G$}

\newcommand{\msun}{$\rm M_\sun$}
\newcommand{\cli}{Class \uppercase\expandafter{\romannumeral1}}
\newcommand{\clii}{Class \uppercase\expandafter{\romannumeral2}}
\newcommand{\cliii}{Class \uppercase\expandafter{\romannumeral3}}
\newcommand{\clitoii}{Class \uppercase\expandafter{\romannumeral1}/\uppercase\expandafter{\romannumeral2}}
\newcommand{\cliitoiii}{Class \uppercase\expandafter{\romannumeral2}/\uppercase\expandafter{\romannumeral3}}
\newcommand{\uco}{$^{12}$CO}
\newcommand{\lco}{$^{13}$CO}
\newcommand{\ltco}{C$^{18}$O}

\newcommand{\eplx}{$ \sigma_\varpi $}
\newcommand{\epmra}{$ \sigma_{\mu_{\alpha^{*}}} $}
\newcommand{\epmdec}{$ \sigma_{\mu_{\delta}} $}

\graphicspath{{./}}

\begin{document}

\title{Star Formation in the \ion{H}{2} Region Sh 2-205: 3D Morphology and Kinematics from Young Stars and Molecular Gas}

\author[0009-0000-5512-9109,gname=Yiwei,sname=Dong]{Yiwei Dong}
\affiliation{Purple Mountain Observatory, Chinese Academy of Sciences, Nanjing 210023, People's Republic of China}
\affiliation{School of Astronomy and Space Science, University of Science and Technology of China, Hefei 230026, People's Republic of China}
\email{dongyw@pmo.ac.cn}

\author[0000-0002-6820-198X,gname=Chaojie,sname=Hao]{Chaojie Hao}
\affiliation{Purple Mountain Observatory, Chinese Academy of Sciences, Nanjing 210023, People's Republic of China}
\email{cjhao@pmo.ac.cn}  

\author[0000-0001-5602-3306,gname=Ye,sname=Xu]{Ye Xu}
\affiliation{Purple Mountain Observatory, Chinese Academy of Sciences, Nanjing 210023, People's Republic of China}
\affiliation{School of Astronomy and Space Science, University of Science and Technology of China, Hefei 230026, People's Republic of China}
\affiliation{Xinjiang Astronomical Observatory, Chinese Academy of Sciences, Urumqi, Xinjiang, 830011, People's Republic of China}
\email[show]{xuye@pmo.ac.cn}  
\correspondingauthor{Ye Xu}

\author[0000-0001-7526-0120,gname=Yingjie,sname=Li]{Yingjie Li}
\affiliation{Purple Mountain Observatory, Chinese Academy of Sciences, Nanjing 210023, People's Republic of China}
\email{liyj@pmo.ac.cn}

\author[0009-0006-0392-6345,gname=Zehao,sname=Lin]{Zehao Lin}
\affiliation{Purple Mountain Observatory, Chinese Academy of Sciences, Nanjing 210023, People's Republic of China}
\email{linzh@pmo.ac.cn}  

\author[0009-0001-9837-9455,gname=Dejian,sname=Liu]{Dejian Liu}
\affiliation{College of Science, China Three Gorges University, Yichang 443000, People's Republic of China}
\email{liudejian@ctgu.edu.cn} 

\author[0000-0002-3904-1622,gname=Yan,sname=Sun]{Yan Sun}
\affiliation{Purple Mountain Observatory, Chinese Academy of Sciences, Nanjing 210023, People's Republic of China}
\email{yansun@pmo.ac.cn}

\author[gname=Longhui,sname=Yang]{Longhui Yang}
\affiliation{Guangxi Key Laboratory for Relativistic Astrophysics, School of Physical Science and Technology, Guangxi University, Nanning 530004, People's Republic of China}
\affiliation{Purple Mountain Observatory, Chinese Academy of Sciences, Nanjing 210023, People's Republic of China}
\email{yang_longhui@st.gxu.edu.cn}

\begin{abstract}
Using \gaia~astrometry of young stars combined with CO observations, we present the first systematic three-dimensional (3D) analysis of the structure, kinematics, and evolutionary history of the star-forming regions in the environs of the \ion{H}{2} region Sh 2-205 (S205). S205 exhibits a complex morphology and coherent expansion on both global and subregional scales. 
We identify several O9--B1 stars and a 0.56 Myr old pulsar that are likely associated with the region. 
A momentum estimate suggests that feedback from these objects may account for the observed overall expansion.
Trace-back analysis of the expansion, combined with color-magnitude diagram fitting {for} young star clusters, indicates at least two episodes of star formation. These results reveal a complex star-formation history of S205 and provide new insights into its 3D evolution.
\end{abstract}


\section{Introduction}
Molecular clouds (MCs) are {the} fundamental component of the interstellar medium and the birthplaces of stars. During the formation of stars and clusters, stellar feedback, {via} photoionization, stellar winds, and supernova explosions, injects energy and momentum into the surrounding interstellar medium. This process continuously shapes the morphology and kinematics of the parent cloud and significantly influences its evolution and subsequent star formation \citep[e.g.,][]{krumholz2014,Chevance2020}.
Revealing the full three-dimensional (3D) spatial and kinematic structure of MCs is therefore essential for understanding their evolution and star-formation history. However, two-dimensional (2D) observations (e.g., molecular line data and infrared imaging) cannot directly determine cloud distances or proper motions, thus hindering {the investigation of} their intrinsic 3D configuration.

Young stellar objects (YSOs) and young cluster members may remain closely associated with their parent clouds. Their spatial and kinematic properties potentially preserve the imprint of the natal cloud and can therefore be used to infer its distance and motion \citep[e.g.,][]{lada1987,Tobin2009,Hacar2016,Ducourant2017,orion2021,gupta2022,liudj2024,dong2024,YangL2025}. With the high-precision parallaxes and proper motions from \gaia~\citep{gaia2016}, it has become feasible to resolve the large-scale 3D structure and kinematics of MCs.
Recent \gaia-based studies illustrate this potential. In Orion, the inferred 3D shape and coherent radial motion suggest {that} this region may have been shaped by past supernova activity \citep[e.g.,][]{orion2018,orion2021}. In the Scorpius-Centaurus OB association, 3D studies suggest a key role of stellar feedback in driving star formation \citep[e.g.,][]{posch2023,Posch2025,MiretRoig2025,Grossschedl2026,Hutschenreuter2026}.
These results underscore the importance of 3D analyses for 
{understanding} the evolution of star-forming regions.

Various physical processes, such as \ion{H}{2} region expansion, supernova explosions, and cloud-cloud collisions, can drive the evolution of star-forming complexes. Systematic 3D studies of these environments provide critical observational constraints on star formation theories. In the supernova remnant region {of} Canis Major (CMa), for example, we have revealed a slowly expanding molecular shell in 3D for the first time. {It has} a dynamical age of $\sim$4 Myr and {shows indications of} multiple supernova events \citep{dong2024}. For \ion{H}{2} regions, while 2D kinematic investigations have been conducted on the Orion Nebula \citep{Pabst2019,Pabst2020}, the existing 3D analyses mainly focus on the larger-scale environments around this \ion{H}{2} region \citep{orion2021}. {The} full 3D {structure and kinematics} of star-forming regions in the environment of {an} \ion{H}{2} region remains poorly understood, motivating us to extend this approach to such environments.

This study focuses on the \ion{H}{2} region Sh 2‑205 \citep[S205;][]{Sharpless_1959}, a $\sim$70--90 pc nebula in the Camelopardalis (Cam) OB1 association \citep[e.g.,][]{straizys2007,straizys2008}. S205 contains multiple substructures, including three H$\alpha$ bubbles (Sh 149.25$-$0.0, Sh 148.83$-$0.67, and LBN 148.11$-$0.45) and one \ion{H}{1} shell. Sh 149.25$-$0.0, Sh 148.83$-$0.67, and the \ion{H}{1} shell were identified and studied in detail by \citet{Romero2008,Romero2009}, and LBN 148.11$-$0.45 was originally cataloged by \citet{Lynds1965} and also analyzed in \citet{Romero2008,Romero2009}. While the bubble expansion may have triggered the local star formation {activity} \citep{Romero2009}, the region's 3D structure and dynamics remain poorly constrained.

Outstanding issues regarding S205 include: (1) uncertain distance estimates \citep[{ranging} from 0.5 to 1.1 kpc; e.g.,][]{FichBlitz1984,Avedisova1984,foster2006,Hou2014,foster2015,Straizys2016,yan2021}, obscuring an accurate 3D physical picture; (2) a lack of detailed 3D kinematic studies, leaving the triggering mechanism and evolutionary history poorly constrained; (3) {the} unclear identity of the main driving source(s) \citep{Sharpless_1959,Avedisova1984,foster2006,Romero2008}; and (4) the true connection of the \ion{H}{2} region to the existing multiple young stellar clusters \citep[e.g.,][]{Romero2008,Romero2009,Hunt2024}.

Using astrometric parameters of young stars from \gaia~Data Release 3 \citep[DR3;][]{gaiadr3}, together with CO data from the Milky Way Imaging Scroll Painting (MWISP) survey \citep{yangji2026}, we derive a reliable distance for S205 along with its internal distance variation. Then, we construct {the} 3D morphology and kinematics {of S205}. By examining massive stars and possible remnants, we explore potential drivers of cloud evolution. Further 3D kinematic analysis reveals its star formation history and dynamical evolution. Using the age distribution of young clusters, we trace the sequence of star formation events in the region. Our work provides a new 3D perspective {on} star formation and evolution in the \ion{H}{2} region environment.

The paper is structured as follows: Section~\ref{sec:data} describes the CO data, {the} YSO sample, and {the} young cluster members. Methods are detailed in Section~\ref{sec:methods}. Section~\ref{sec:3d_results} presents the 3D morphology and kinematics through subregional analysis. In Section~\ref{sec:sources}, we explore the origin of the potential drivers of S205. Section~\ref{section:history} constructs the star formation history using {the} dynamical timescale and ages of star clusters. Section~\ref{sec:summary} provides a summary of {this} work.

\section{Data}\label{sec:data}
\subsection{Molecular Clouds}\label{sec:cloud}

\begin{deluxetable*}{chccccrrrhc}
	\tablecaption{\uco~Molecular Clouds in the S205 Region
	\label{tab:mcs}}
	\centering
	\tablehead{
	\colhead{No.} &\nocolhead{idx}&\colhead{Name} &\colhead{$ l\rm_{min} $}&\colhead{$ l\rm_{max} $}&\colhead{$ b\rm_{min} $}&\colhead{$ b\rm_{max} $}&\colhead{ $v\rm_{LSR, min}$}&\colhead{ $v\rm_{LSR, max}$}&\nocolhead{$ v\rm_{rms} $}&\colhead{Subregion} \\
	\colhead{} &\colhead{}&\colhead{} &  \colhead{($\degree$)}&\colhead{($\degree$)}&\colhead{($\degree$)}&\colhead{($\degree$)}& \colhead{($\rm km\,s^{-1}$)}& \colhead{($\rm km\,s^{-1}$)} &
	\nocolhead{($\rm km\,s^{-1}$)}&\colhead{}
	}
	\decimalcolnumbers
	\startdata
	1&48493&MWISP G147.590$-$01.153
	&146.84&148.31& $-$1.70& $-$0.58& $-$5.56&  3.65&  0.87&A\\
	2&35825&MWISP G148.153$-$00.241
	&146.89&149.01& $-$1.22&  0.88&$-$15.72&  2.06&  2.65&B\\
	3&37658&MWISP G149.528$-$01.081
	&148.90&149.97& $-$1.78& $-$0.16&$-$14.13& $-$0.95&  1.82&C/E\\
	4&36796&MWISP G150.216$-$01.278
	&150.00&150.50& $-$1.61& $-$1.07&$-$14.76& $-$3.02&  2.74&D\\
	5&37828&MWISP G149.139$-$02.111
	&148.92&149.47& $-$2.42& $-$1.88&$-$14.13& $-$3.49&  1.12&F\\
	\enddata
	\tablecomments{
	Column (1): cloud index. 
	Column (2): cloud name given by the Galactic coordinates of the cloud centroid. 
	Columns (3--8): the Galactic longitude, latitude, and LSR velocity ranges of MCs derived from \uco~observations. 
	Column (9): subregion to which each cloud is assigned.}
\end{deluxetable*}

%
This study utilizes {the} \uco~and \lco~(1--0) line data from the MWISP DR1 \citep{yangji2026}, observed using the
Purple Mountain Observatory (PMO) 13.7 m millimeter-wavelength telescope with a beam size of $\sim$50\arcsec~at 115 GHz. The median root-mean-square (RMS) noise levels for the \uco~and \lco~lines are 0.48 and 0.26 K, respectively.

Along the line of sight toward S205, there are several layers of molecular gas with clearly separated local standard of rest (LSR) velocities \citep{Digel1996, Du_2017}. This region is part of the Cam OB1 layer \citep{straizys2007, straizys2008}, with its molecular gas distribution shown in Appendix~\ref{app:auxiliary} Figure~\ref{fig:allv}. We identify $^{12}$CO clouds and their $^{13}$CO dense structures using the DBSCAN-based \citep{Ester_1996} algorithm from \citet{Yan_2020}, {which has been} validated across multiple Galactic regions \citep[e.g.,][]{Yan_2020, yan2021, Sun_2021,Yuan_2022,Dong2023}. Five major $^{12}$CO clouds containing $^{13}$CO structures are identified, with {their} spatial and velocity ranges listed in Table~\ref{tab:mcs}. Faint peripheral clouds are not considered in our analysis.

\subsection{{Young Stars}} \label{sec:young_stars}
\subsubsection{{YSOs}}\label{sec:ysos}

\begin{deluxetable*}{lllcc}
	\tablecaption{Summary of YSOs from the Literature \label{tab:samples}}
	\tablehead{\colhead{Reference}	&\colhead{Provided Designation}	&\colhead{Coverage}&\colhead{$N\rm_{YSO} $} & \colhead{$N_{Gaia} $}
	}
	\colnumbers
	\startdata
	\citet{marton2016}	&AllWISE& All-sky 
	 &130		&74\\
	\citet{Straizys2016}	&AllWISE& $l\sim148\degree,\,b\sim-0\degree $~($\sim$$ 1.5\degree \times 1\degree$)
	 &88		&44\\
	\citet{w20}	&2MASS, AllWISE& $65\degree\lesssim l\lesssim265\degree,\,|b|\lesssim 3\degree $
	&  276&172\\	
	\citet{Wilson2023}	&\gaia~EDR3& $20\degree<l<220\degree,\,|b|< 4\degree $&  57	&57\\
	\cite{z23}	&\gaia~DR3, AllWISE& All-sky   &194	&194\\	
	\citet{mar23}	&\gaia~DR3& 
	Nearby star-forming regions
	&8		&8\\
	\citet{zhang2024}	&\gaia~DR3
	& $30\degree\lesssim l\lesssim220\degree,|b|\lesssim 20\degree$
	&7	&7\\
	\enddata
	\tablecomments{
	Column (2): infrared and/or \gaia~IDs given by the catalogs.
	Column (3): the covered Galactic area.
	Column (4): the number of YSOs toward the S205 region.
	Column (5): the number of YSOs after cross-matching with \gaia~DR3.}
\end{deluxetable*}

YSOs can serve as tracers of  their parent cloud's distance and kinematics \citep{gutermuth2011, gh21, orion2021}. Within the S205 region, we compile published YSO catalogs, retaining only sources at Class II evolutionary stages or earlier. The sample comprises 760 entries drawn from seven references {listed in Table~\ref{tab:samples}, including six large-scale surveys \citep{marton2016,w20,Wilson2023,z23,mar23,zhang2024} and one local survey toward S205 \citep{Straizys2016}}. In the large-scale YSO surveys, potential contamination (e.g., from asymptotic giant branch (AGB) stars) has been discussed and filtered {out}.

The \gaia~mission \citep{gaia2016, gaiadr3} provides precise astrometry for 1.5 billion sources {from} 34 months of observations, including parallaxes ($\varpi$), proper motions ($\mu_{\alpha^*}$, $\mu_{\delta}$), {and} radial velocities for 33 million stars.
We cross-match YSOs with {the} \gaia~DR3 dataset via the \gaia~$\texttt{source\_id}$, \emph{allwise\_best\_neighbour} and \emph{tmass\_psc\_xsc\_best\_neighbour} tables, directly using the corresponding IDs as listed in Column 2 of Table~\ref{tab:samples}. Nine ambiguous matches {that resolved} to two \gaia~sources are excluded, yielding 556 reliable matches.

Following the recommendations of \citet{Fabricius2021} and \citet{Gaia2021}, 
sources with $\texttt{ipd\_gof\_harmonic\_amplitude}>0.1$ combined with $\texttt{ruwe} > 1.4$ are identified as non-single stars or sources with poorly
modeled astrometric solutions.
These sources are excluded from the sample, leaving 524 YSO entries. We further apply a cut of $\texttt{parallax\_over\_error} > 3$ to {remove} sources of poor astrometric quality, yielding 319 YSO entries.
Through inspection, 62 YSOs are detected by multiple studies and are classified consistently, thus appearing as repeated entries. {The remaining} 156 YSO entries are unique.
After removing duplicates, the sample contains 218 YSOs.

The parallax and distance distributions of YSOs (Appendix~\ref{app:auxiliary} Figure~\ref{fig:plx}) show clear and narrow {peaks} around 0.9 mas and 1.1 kpc, respectively, broadly consistent with the commonly adopted distance of $\sim$1 kpc \citep{Romero2008,Romero2009}. Guided by these distributions, we adopt a parallax range of 0.7--1.5 mas to define a preliminary YSO sample, reducing foreground and background contamination and leaving 107 YSOs. Sources beyond three standard deviations from the mean parallax or proper motions are iteratively removed until convergence (Appendix~\ref{app:auxiliary} Figure~\ref{fig:plx_pm}), ultimately yielding 78 high-quality YSOs likely associated with S205. Finally, we verify that no potential AGB contaminants remain in the sample by examining the \gaia\ color--magnitude diagram (CMD).

\subsubsection{Member Stars of Young Star Clusters}\label{sec:ocs}

The S205 region contains several young clusters with reported ages of~$\lesssim$10 Myr. 
Clusters in this age range are generally expected to remain associated with their parent molecular clouds, as suggested by both observational and numerical studies \citep[e.g.,][]{Kim2018,Grasha2018,Peltonen2023}. Nevertheless, the close kinematic association between young cluster members and their parent molecular clouds still needs to be verified for individual star-forming regions. We adopt the comprehensive \gaia~DR3 open cluster catalog of \citet{Hunt2024}. {The} parameters and member star lists of four young clusters in the S205 region are retrieved from \citet{Hunt2024yCat} via the VizieR database.
Their properties are summarized in Table~\ref{tab:ocs}.

Our final catalog of young stars is a combination of the selected YSOs and cluster members. 
We identify 21 duplicates between the two samples. This is a confirmation of the robust membership of these 21 stars. To avoid duplicates in our final catalog, we merge the YSO and cluster samples, while keeping flags in our catalog that indicate the original cluster membership and/or the YSO nature of each star.
The resulting young star sample {is} presented in Table~\ref{tab:ysos}.

In Figure~\ref{fig:rgb} (a)--(c), member stars of young clusters and YSOs show consistent spatial, parallax, and proper motion distributions, supporting {the} treatment {of them} as a unified young star sample for analyzing the 3D structure and kinematics of S205. This coherence aligns with previous studies of other star-forming regions \citep{Tobin2009,Hacar2016,orion2021,Roychowdhury2024, Roychowdhury2025}, {indicating} that young stellar populations share coherent spatial and kinematic distributions. Histograms of the astrometric parameters of young stars in the final catalog are presented in Figure~\ref{fig:a4}.

\begin{deluxetable*}{lcccccccccccc}
	\tablecaption{{Young Star Clusters in the S205 Region}
	\label{tab:ocs}}
	\centering
	\tablehead{
	\colhead{Name} &\colhead{$N_*$} &\colhead{$ l $}&\colhead{$ b $}&\colhead{Dist.}&\colhead{\pmra}&\colhead{\pmdec}&\colhead{$X$}&\colhead{$Y$}&\colhead{$Z$}&\colhead{Age}&\colhead{Mass}&\colhead{Subregion}\\
	\colhead{} & \colhead{} &  \colhead{($\degree$)}&\colhead{($\degree$)}&\colhead{(pc)}& \colhead{($\rm mas\,yr^{-1}$)}& \colhead{($\rm mas\,yr^{-1}$)}&\colhead{(pc)}&\colhead{(pc)} &
	\colhead{(pc)}&\colhead{(Myr)}&\colhead{(\msun)}&\colhead{}
	}
	\decimalcolnumbers
	\startdata
	NGC 1444&104&$148.1$&$-1.3$&$1103$&$-0.8$&$-1.8$&$-9058$&$583$&$-2$&$7.4^{+3.5}_{-2.3}$&$407\pm25$&A\\
	CWNU 1042&49&$148.3$&$-0.4$&$1124$&$-0.4$&$-1.5$&$-9078$&$591$&$15$&$5.9^{+2.7}_{-1.9}$&$163\pm48$&B\\
	UBC 51&144&$149.2$&$-0.4$&$1052$&$-0.3$&$-1.4$&$-9026$&$538$&$16$&$6.0^{+2.3}_{-1.9}$&$342\pm43$&C\\
	HSC 1193&14&$150.0$&$-1.1$&$1042$&$-0.02$&$-2.0$&$-9024$&$521$&$3$&$7.8^{+3.9}_{-3.8}$&$82\pm22$&D\\
	\enddata
	\tablecomments{Column-wise: cluster name, number of member stars, Galactic longitude and latitude ($l, b$), distance, proper motions (\pmra, \pmdec), Galactic Cartesian coordinates ($X, Y, Z$), age, mass, and the associated subregion of each cluster. Properties in Columns (1)--(12) are adopted from \citet{Hunt2024yCat}.}
\end{deluxetable*}

\begin{deluxetable*}{ccccccccccccc}
	\tabletypesize{\footnotesize}
	\tablecaption{Catalog of Young Stars in the S205 Region	\label{tab:ysos}}
	\tablehead{\colhead{\gaia~DR3}	&\colhead{$ l $}&\colhead{$ b $}&\colhead{\plx} & \colhead{\eplx}&\colhead{\pmra}&\colhead{\epmra}&\colhead{\pmdec}&\colhead{\epmdec}&\colhead{YSO Flag}&\colhead{YSO Ref.}&\colhead{\citetalias{Hunt2024} Cluster}&\colhead{Subregion}\\
	\colhead{\texttt{source\_id}}	&\colhead{(\degree)}&\colhead{(\degree)}&\colhead{(mas)} & \colhead{(mas)}&\colhead{(\masyr)}&\colhead{(\masyr)}&\colhead{(\masyr)}&\colhead{(\masyr)}&\colhead{}&\colhead{}&\colhead{}&\colhead{}}
	\decimalcolnumbers	
	\startdata
	444035507433873280&147.09&$-1.44$&0.70&0.12&$-0.70$&0.12&$-1.13$&0.12&YSO&\citet{marton2016}&\nodata&A\\
	252050297504923008&148.22&$-0.41$&0.86&0.06&$-0.46$&0.07&$-1.73$&0.07&YSO&\citet{Wilson2023}&CWNU\_1042&B\\
	251756242571599104&149.31&$-0.56$&0.87&0.20&$-0.14$&0.19&$-1.30$&0.16&\nodata&\nodata&UBC\_51&C\\
	250860935161004288&150.20&$-1.10$&0.91&0.16&$-0.05$&0.17&$-2.10$&0.15&\nodata&\nodata&HSC\_1193&D\\
	250897901941962752&149.76&$-1.42$&0.96&0.10&$-0.67$&0.10&$-2.35$&0.09&YSO&\citet{Wilson2023}&\nodata&E\\
	251339703754539776&149.07&$-2.01$&1.12&0.19&$-0.93$&0.20&$-2.05$&0.19&YSO&\citet{marton2016}&\nodata&F\\
	\nodata&   &  &  &  &  &  &  &&  \\
	\enddata
	\tablecomments{Columns (1--9): \gaia~source identifier and astrometric parameters of young stars. Column (10): YSO flag. Column (11): reference(s) identifying the source as a YSO. Column (12): associated cluster from \citet{Hunt2024}. Column (13): subregion to which each source is assigned. This table is available in its entirety in machine-readable form.}	
\end{deluxetable*}

\section{Methods\label{sec:methods}}

\subsection{3D Positions and Velocities of star-forming regions in S205} \label{sec:6d_parameter_determination}

There are several star-forming subregions and associated young stellar associations in the environs of the \ion{H}{2} region S205. To investigate the internal structure and kinematics of S205, we derive the 3D positions and velocities of these stellar associations in a local Cartesian coordinate system. The analysis proceeds as follows.

First, we derive the average properties for each subregion by calculating the mean astrometric parameters ($l$, $b$, \plx, $d$, \pmra, \pmdec) and the mean radial velocities (RVs). 
The mean astrometric parameters are obtained by averaging the observed parameters 
of the young stars.
The corresponding uncertainties are estimated using 
a Monte Carlo sampling approach, i.e., by calculating the standard deviations of 
the simulated mean values obtained from Gaussian realizations based on the 
\gaia~measurements and their associated uncertainties.
The same procedure is adopted throughout the following analysis.
Distances are derived from the inverse of the mean parallax ($d=1/\varpi$). 

The RVs of young stars in the S205 region with small RV uncertainties ($<5$ \kms) are roughly concordant with those of molecular gas within the uncertainties (see Appendix~\ref{app:rvs}). In individual star-forming subregions, however, only a few or no young stars have small RV uncertainties. The current data only tentatively {suggest} that the stellar and gas RVs in the subregions lie within similar ranges. Given the uncertainties, this conclusion remains ambiguous. As shown in Appendix Table~\ref{tab:subregion_rvs} and Figure~\ref{fig:subregion_rvs}, the stellar RV subsamples in some subregions appear to deviate from the gas RVs or show sensitivity to the adopted RV uncertainty thresholds. Therefore, given the limited number of young stars with available high-precision RV measurements, the close kinematic connection between young stars and molecular gas in S205 requires confirmation {by} additional stellar RV measurements.

Here, we adopt the mean RV of the gas within each subregion to represent its RV. This approximation provides only a rough estimate, especially {for} the relatively old clusters. Specifically, we derive the main-beam brightness temperature ($T\rm_{MB}$)-weighted mean \vlsr~of {the} \lco~observations, since \lco~better traces dense structures within clouds. The \lco~velocity dispersion is adopted as the corresponding uncertainty.
For consistency with the heliocentric astrometric parameters, the adopted gas \vlsr~values 
are converted to heliocentric RVs (\vhel), using the conventional definition of the 
LSR \citep{Gordon1976}.
Finally, the astrometric parameters and RV of the reference center
are obtained by averaging the corresponding parameters of all subregions. The above mean parameters are listed in Table~\ref{tab:basic_para}.

\begin{figure*}[t!]
	\centering
	\includegraphics[width=0.95\textwidth]{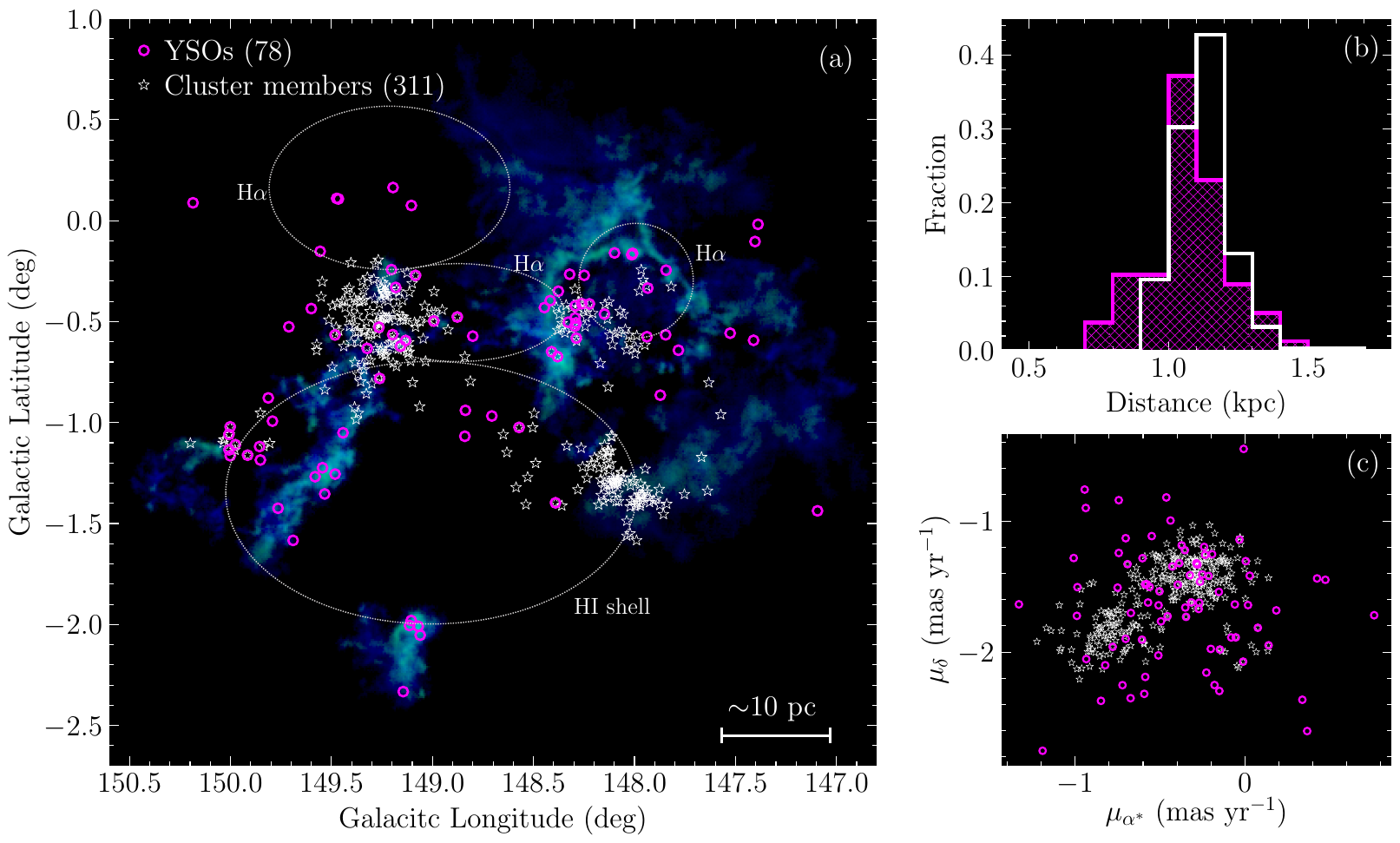}
	\caption{Distributions of young stars and molecular gas. 
	(a) $l$--$b$ distribution of YSOs (magenta circles) and young cluster members (white star symbols) overlaid on the \uco~(blue) and \lco~(green) intensity maps. The big circles are {three} H$\alpha$ bubbles and {an} \ion{H}{1} shell depicted in Figure 1 of \citet{Romero2009}.
	(b) Histograms of distances of YSOs (magenta) and young cluster members (white). 
	(c) Proper motion distributions of YSOs and young cluster members.\label{fig:rgb}}
\end{figure*}

Second, following the method of \citet{Reid_2019}, the mean 
parameters ($l$, $b$, $d$, \pmra, \pmdec, \vhel) are transformed into 3D 
positions ($x$, $y$, $z$) and 3D velocities ($v_x$, $v_y$, $v_z$). In this paper, we adopt the Heliocentric Galactic Cartesian coordinate system following the convention commonly used in \gaia-based studies. The origin is located at the position of the Sun. The $x$-axis is positive toward the Galactic center, the $y$-axis is positive in the direction of Galactic rotation, and the $z$-axis is positive toward the north Galactic pole. The adopted Galactic parameters and the used standard solar motions are 
$R_0=8.15\pm0.15$ kpc, $\Theta_0=236\pm7$ \kms, and 
$(U_0, V_0, W_0)=(10.6\pm1.2, 10.7\pm6.0, 7.6\pm0.7)$ \kms~\citep{Reid_2019}. The uncertainties of the 3D positions and velocities are estimated 
using Monte Carlo simulations. The resulting 3D positions and velocities are listed in Table~\ref{tab:3d_rel}.

Third, we define a local Cartesian coordinate system 
($o_{\rm c}$-$x_{\rm c}y_{\rm c}z_{\rm c}$) with its origin at the reference center of S205
and its axes aligned with those of the Heliocentric Galactic Cartesian coordinate system.
The 3D positions, $\boldsymbol{r}=(x_{\rm c}, y_{\rm c}, z_{\rm c})$, and velocities,
$\boldsymbol{v}=(v_{x_{\rm c}},v_{y_{\rm c}},v_{z_{\rm c}})$, 
of each subregion are obtained by subtracting the position and velocity of the 
reference center from the corresponding values of the subregion
in the Heliocentric Galactic Cartesian coordinate system, respectively (Table~\ref{tab:3d_rel}; Figure~\ref{fig:3d_project}). 

Following \citet{rivera2015} {and} \citet{Grossschedl2026}, we compute the radial component 
of {the} 3D velocity of each subregion relative to the reference center as 
$v_r=\hat{\boldsymbol{r}}\cdot \boldsymbol{v}$, 
where $\hat{\boldsymbol{r}}=\boldsymbol{r}/\vert\boldsymbol{r}\vert$ is the unit position 
vector and $\vert\boldsymbol{r}\vert$ is the 3D distance from the reference center. The mean radial velocity, $\overline{v_r}$, serves as an indicator of expansion or 
contraction of the S205 region, with positive values corresponding to outward expansion (see Table~\ref{tab:3d_rel}).

\subsection{Internal Kinematics of Star Clusters in S205 \label{sec:method_3.2.2}}

The young stars in S205 {belong to} six star clusters. Since coherent expansion is commonly observed in young clusters \citep[e.g.,][]{Kuhn2019,Della2024}, we analyze the internal kinematics of {the} star clusters in S205. Cluster members (i.e., young stars) with parallaxes or proper motions deviating from the cluster mean by more than $3\sigma$ are excluded. Proper motions of cluster members are transformed from equatorial (\pmra, \pmdec) to Galactic coordinates ($\mu_l$, $\mu_b$). To mitigate the impact of parallax uncertainties of individual stars, we assume that all stars of a cluster lie at the cluster mean distance to convert proper motions into tangential velocities ($v_l$, $v_b$). 

We compute the 2D position and velocity vectors of the $k$-th young star, $\boldsymbol r_k=(l_k-l, b_k-b)$ and $\boldsymbol v_k=(v_{l,k}-v_{l}, v_{b,k}-v_{b})$, where $(l,b)$ and $(v_{l},v_{b})$ denote the mean position and tangential velocity components of the cluster, respectively, as listed in Table~\ref{tab:basic_para}. The radial velocity component is $v_{r, k}=\hat{\boldsymbol{r_k}}\cdot \boldsymbol{v_k}$, where $\hat{\boldsymbol{r_k}}=\boldsymbol{r_k}/\vert\boldsymbol{r_k}\vert$. Finally, the average value of all cluster members, $\overline{v_{r,k}}$, is used to characterize cluster expansion or contraction, with positive values indicating expansion. We also calculate the ratio $\overline{v_{r,k}}/\sigma_{v_{r,k}}$, where $\sigma_{v_{r,k}}$ is the radial velocity dispersion, to quantify the relative importance of ordered and disordered radial motions \citep[e.g.,][]{Della2024}. The resulting $\overline{v_{r,k}}$ and $\overline{v_{r,k}}/\sigma_{v_{r,k}}$ values are shown in Figure~\ref{fig:subregion_expansion}.

\subsection{Age Determination of Star Clusters in S205}\label{sec:stellar_age_determination}
Following the method described in \citet{Liu_pang2019} and \citet{hao2022}, we use the \gaia~photometry ($G$, $G_{\rm BP}$, $G_{\rm RP}$) to construct the CMDs of young stars and derive the cluster ages through isochrone fitting (Figure~\ref{fig:regions}). The theoretical isochrones are taken from the PARSEC library \citep{parsec_isochrones}, which has been updated for the \gaia~passbands \citep{Evans2018}. The mean squared distance $\overline{d^{2}}$ between the young stars and the theoretical isochrones is defined as
\begin{equation} 
\centering
\overline{d^{2}} = \sum_{k=1}^{N_*} | \ {\emph{x}}_{k} - {\emph{x}}_{k, nn} \ |^{2} / N_*,
\end{equation}
where $N_*$ is the number of young stars, ${\emph{x}}_{k} = [{\it G}_{k} + \Delta_{\rm G} + {\it A}_{\rm G}, ({\it G}_{\rm BP} - {\it G}_{\rm RP})_{\it k} + {\it E}({\it G}_{\rm BP} - {\it G}_{\rm RP})]$ denotes the CMD position of the $k$th young star, and ${\emph{x}}_{k, nn}$ is the nearest neighboring point to this star on the isochrone. $\Delta_{\rm G}$ is the distance modulus, ${\it A}_{\rm G}$ is the extinction, and ${\it E}({\it G}_{\rm BP} - {\it G}_{\rm RP})$ is the reddening. During the fitting procedure, we vary $A_{\rm G}$ from 0 to 4 mag in steps of 0.02 and adopt $E(G_{\rm BP}-G_{\rm RP}) = 0.50\,A_{\rm G}$, considering the extinction curve of the Milky Way \citep{Cardelli1989,ODonnell1994} and \gaia~extinction relations \citep{Andrae2018}. To ensure robust fits, we restrict the sample to young stars with ${\rm G}<18$, which retains over 60\% of the targets. The cluster ages are listed in Table~\ref{tab:basic_para} as basic properties of the subregions.

\section{Morphology and Motions of S205}\label{sec:3d_results}

\begin{figure}[t!]
	\centering
	\includegraphics[width=0.95\textwidth]{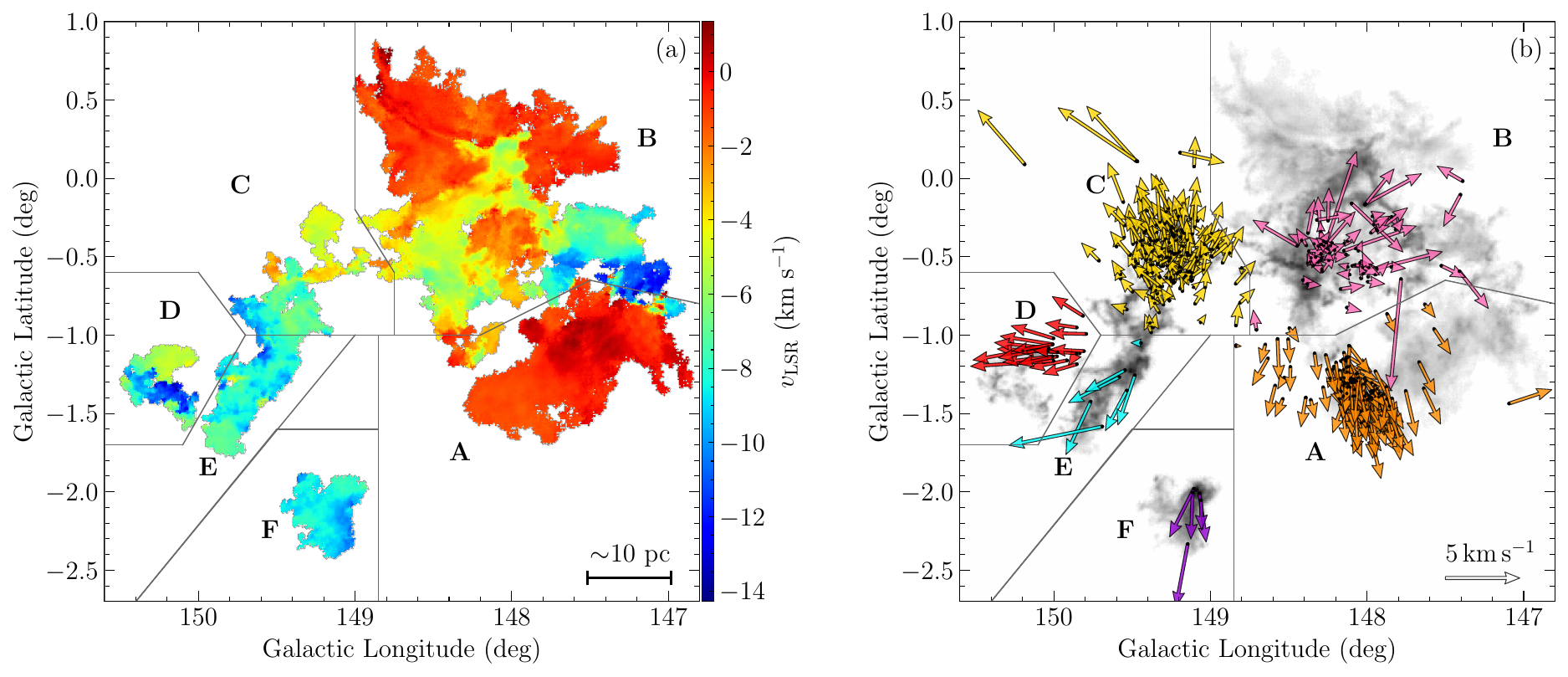}
	\caption{(a) LSR velocity distribution map of \uco~emission. Gray solid lines mark the boundaries of six identified subregions (A--F). (b) Locations and residual tangential motions of young stars. Black dots denote locations of young stars, and arrows depict their residual tangential motions after subtracting the bulk tangential motion of the region. Here, the unit of \masyr~{is} converted to \kms~using the parallaxes of young stars. Subregions A--F are colored orange, pink, yellow, red, cyan, and purple, respectively. \label{fig:2dmotion}}
\end{figure}

\subsection{Division of Subregions}  \label{sec:6d_subregion} 

As shown in Figure~\ref{fig:rgb}~(a), S205 {contains} several substructures, 
including three H$\alpha$ bubbles and one \ion{H}{1} shell \citep{Romero2008,Romero2009}. 
To investigate the internal structure and kinematics,
we decompose S205 into several kinematically coherent subregions.
Both the molecular gas and young stars 
exhibit spatial and kinematic variations across S205, as shown in Figure~\ref{fig:2dmotion}. 
On the one hand,
five clouds appear as distinct structures in position-position-velocity (PPV) space, 
each with its own spatial and velocity characteristics (Figure~\ref{fig:2dmotion} (a); Table~\ref{tab:mcs}). 
On the other hand,
the motions of young stars show systematic differences across the region.
Figure~\ref{fig:2dmotion} (b) presents the residual tangential motions of 
young stars after subtracting the overall mean motion. Here, the proper motions (\pmra, \pmdec) have been converted into tangential velocities ($v_l$, $v_b$).

Considering the PPV structure of the MCs and the bulk motion of young stars, 
we decompose S205 into six subregions (A--F), with boundaries indicated by 
the gray solid lines in Figure~\ref{fig:2dmotion}. 
Four young clusters listed in Table~\ref{tab:ocs} are located within subregions A, B, C, and D, respectively. 
Clouds No. 1, 2, 4, and 5 in Table~\ref{tab:mcs} fall within subregions A, B, D, and F, 
respectively.
Cloud No. 3 is split between subregions C and E due to the 
distinct residual tangential motions of the associated young stars (Figure~\ref{fig:2dmotion} (b)). 
In subregion B, the \uco~gas shows a velocity gradient of several \kms~from north 
to south (Figure~\ref{fig:2dmotion} (a)). 
However, the associated \lco~emission (green in Figure~\ref{fig:rgb} (a)) is concentrated 
mainly in the southern part, where most of the young stars are also located. 
To the east of subregion D ($l\sim150.6\degree$--$152\degree$) lie the \ion{H}{2} regions Sh 2-209, NGC 1491, and IRAS 04000$+$5052. We find that the CO velocities and distances reported for these regions in the literature \citep{Wouterloot1989,foster2015} differ significantly from those of S205. Consequently, despite their apparent proximity on the sky, these regions are unlikely to be physically associated with S205.

\begin{figure}[t!]
	\centering
	\includegraphics[width=0.9\textwidth]{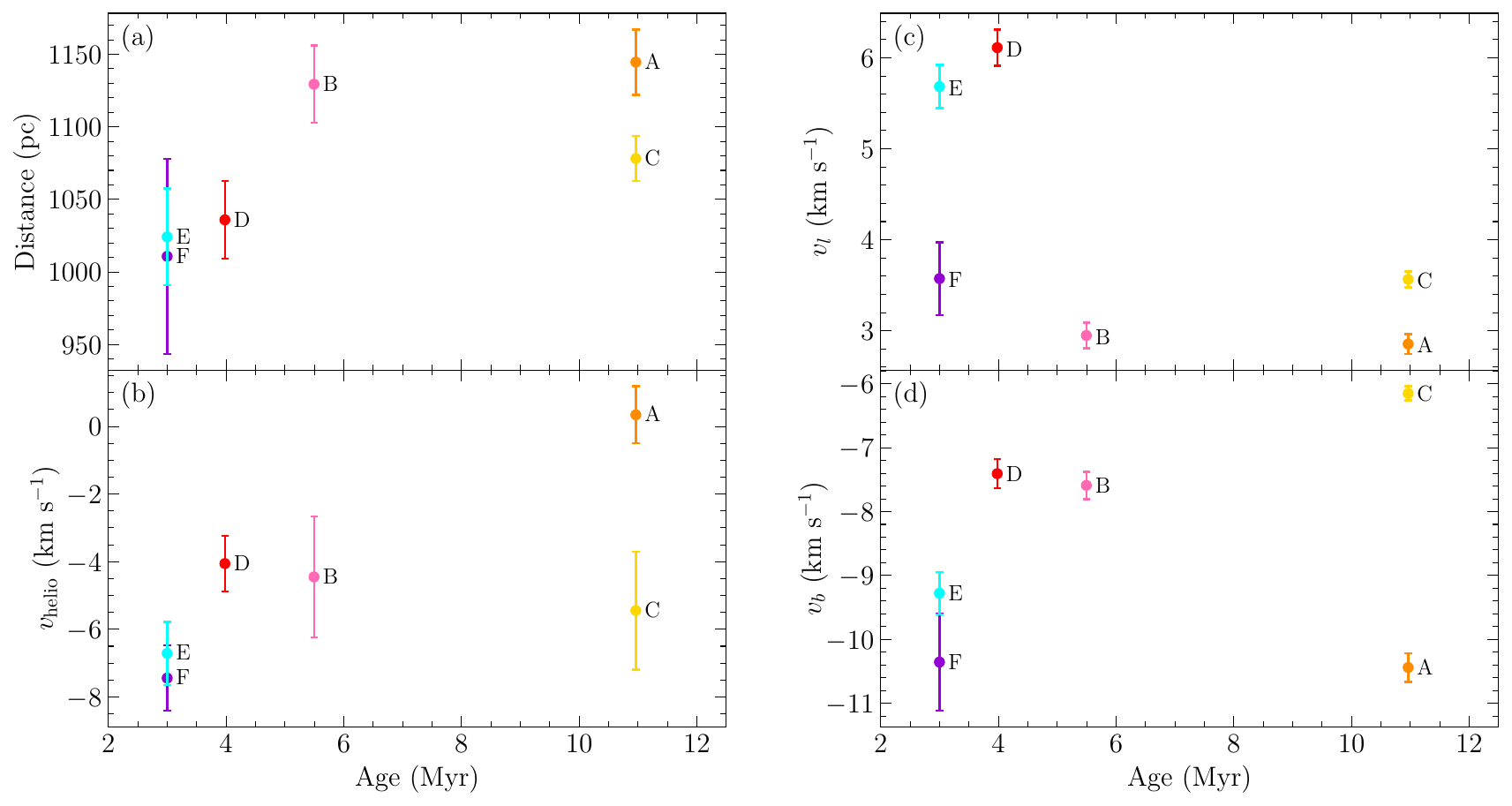}
	\caption{Overview of four average parameters as derived for the six subregions. Panels (a), (b), (c), and (d) show the distances, \vhel, {$v_l$, and $v_b$}, respectively. Error bars represent their respective uncertainties. Colors of subregions match Figure~\ref{fig:2dmotion} (b). 
	The subregions are ordered by age (see Table~\ref{tab:basic_para}). 
\label{fig:variation}}
\end{figure}

Using the method described in Sections~\ref{sec:6d_parameter_determination} {and~\ref{sec:method_3.2.2}}, we 
derive the {average} astrometric parameters and RVs of the six subregions and the reference center. The results are listed in Table~\ref{tab:basic_para}. Figure~\ref{fig:variation} illustrates the distinct distances, heliocentric RVs (\vhel), and {tangential motions ($v_l$ and $v_b$)} of the six subregions.

\begin{deluxetable*}{ccccccccccccccc}
	\tabletypesize{\scriptsize}
	\tablecaption{Basic Properties of Subregions and Reference Center of S205
	\label{tab:basic_para}}
	\tablehead{
		\colhead{Subregion} 
		&\colhead{$N_*$}&\colhead{$ l $}&\colhead{$ b $} & \colhead{\plx} &\colhead{$ d $}&	\colhead{\pmra}  &\colhead{\pmdec}&\colhead{$v_{l}$}&\colhead{$v_{b}$}&\colhead{\vlsr}&\colhead{\vhel} & 	\colhead{$M\rm_{gas}$} &\colhead{$M\rm_{star}$}&\colhead{Age}\\
		\colhead{}&\colhead{}
		&\colhead{(\degree)}&\colhead{(\degree)}&
		\colhead{(mas)} &\colhead{(pc)} & \colhead{(\masyr)} 
		&\colhead{(\masyr)}&
		\colhead{(\kms)}&
		\colhead{(\kms)}&
		\colhead{(\kms)}&
		\colhead{(\kms)}& \colhead{($10^3$ \msun)} &\colhead{(\msun)}&  \colhead{(Myr)} }
	\decimalcolnumbers
	\startdata
	A&107&148.1&$-1.3$&$0.87\pm0.02$&$1145\pm23$&$-0.79\pm0.02$&$-1.83\pm0.02$&$2.85\pm0.11$&$-10.44\pm0.23$&$-0.5$ (0.8)&$0.3\pm0.8$&5.0&$482\pm52$&$11.0^{+4.2}_{-3.0}$\\
	B&72&148.1&$-0.5$&$0.89\pm0.02$&$1129\pm27$&$-0.47\pm0.02$&$-1.45\pm0.02$&$2.95\pm0.14$&$-7.59\pm0.21$&$-5.2$ (1.8)&$-4.5\pm1.8$&21.9&$214\pm40$&$5.5^{+2.0}_{-1.5}$\\
	C&158&149.3&$-0.5$&$0.93\pm0.01$&$1078\pm15$&$-0.25\pm0.01$&$-1.37\pm0.01$&$3.56\pm0.09$&$-6.15\pm0.11$&$-6.5$ (1.7)&$-5.4\pm1.7$&3.3&$406\pm73$&$11.0^{+4.2}_{-3.0}$\\
	D&19&149.9&$-1.1$&$0.97\pm0.02$&$1036\pm27$&$-0.04\pm0.03$&$-1.96\pm0.02$&$6.11\pm0.20$&$-7.41\pm0.22$&$-5.4$ (0.8)&$-4.1\pm0.8$&0.9&$75\pm29$&$4.0^{+1.4}_{-1.0}$\\
	E&7&149.6&$-1.3$&$0.98\pm0.03$&$1024\pm33$&$-0.34\pm0.03$&$-2.22\pm0.03$&$5.68\pm0.24$&$-9.28\pm0.33$&$-8.0$ (0.9)&$-6.7\pm0.9$&4.2&\nodata&$\lesssim 3$\\
	F&5&149.1&$-2.1$&$0.99\pm0.06$&$1011\pm67$&$-0.78\pm0.07$&$-2.15\pm0.06$&$3.57\pm0.40$&$-10.35\pm0.76$&$-8.7$ (1.0)&$-7.4\pm1.0$&2.5&\nodata&$\lesssim 3$\\
	Center & \nodata & 149.0 & $-1.1$ & $0.94\pm0.01$& $1070\pm15$ & $-0.44\pm0.01$ & $-1.83\pm0.01$& $4.12\pm0.09$ & $-8.54\pm0.15$&$-5.7$ (0.5)&$-4.6\pm0.5$&\nodata&\nodata&\nodata\\
	\enddata
	\tablecomments{Column (1): subregion name. 
	Column (2): number of young stars.
	Columns (3--10): averaged parameters estimated from young stars.
	Column (11): $ T\rm_{MB} $-weighted average \vlsr~of the \lco~observations.
	The velocity dispersion in parentheses {is} estimated as the $ T\rm_{MB} $-weighted standard deviation of the \lco-traced \vlsr, serving as the uncertainty.
	Column (12): heliocentric RV converted from the LSR velocity in Column (11).
	Column (13): molecular mass traced by \uco.
	Column (14): stellar mass obtained from fitting mass functions.
	Column (15): cluster age: A--D from CMD fitting (Figure~\ref{fig:regions}); E and F assigned {an} indicative age of $\sim$3 Myr, considering the typical ages of Class II YSOs \citep{Williams2011}.
	}
\end{deluxetable*}

\begin{deluxetable*}{crrrrrrrrrrrrrch}
	\tabletypesize{\scriptsize}
	\setlength{\tabcolsep}{0.8mm}
	\tablecaption{Summary of 3D Positions and Motions of Subregions
	\label{tab:3d_rel}}
	\tablehead{
		\colhead{Subregion} 	&\colhead{$ x$}&\colhead{$ y $} & \colhead{$ z $} &	\colhead{$v_{x} $}&	\colhead{$v_{y} $}&	\colhead{$v_{z} $}
		&\colhead{$ x\rm_{c}$}&\colhead{$ y\rm_{c} $} & \colhead{$ z\rm_{c}$} &	\colhead{$v_{x\rm_{c}} $}&	\colhead{$v_{y\rm_{c}} $}&	\colhead{$v_{z\rm_{c}} $}&	\colhead{$\vert\boldsymbol{r}\vert$}&\colhead{$v_r$}&\nocolhead{$\overline{v_{r, k}}/\sigma_{v_{r, k}}$}\\ 
		\cmidrule(l{4pt}r{4pt}){2-4}
		\cmidrule(l{4pt}r{4pt}){5-7}
		\cmidrule(l{4pt}r{4pt}){8-10}
		\cmidrule(l{4pt}r{4pt}){11-13}
		\colhead{}&\colhead{}&\colhead{(pc)}&\colhead{}&\colhead{}&\colhead{(\kms)}&\colhead{}&\colhead{}&\colhead{(pc)}&\colhead{}&\colhead{}&\colhead{(\kms)}&\colhead{}&\colhead{(pc)}&\colhead{(\kms)}&\nocolhead{}
	}
	\decimalcolnumbers
	\startdata
	A&$-971.4\pm19.1$&$604.8\pm11.9$&$-25.6\pm0.5$&$-6.5\pm0.8$&$10.5\pm0.4$&$-2.8\pm0.2$&$-53.8\pm22.9$&$53.9\pm14.1$&$-4.7\pm0.6$$\ $&$-4.8\pm0.9$&$3.9\pm0.5$&$-2.0\pm0.3$&$76.3\pm25.7$&$6.3\pm1.1$\\
	B&$-959.1\pm22.6$&$596.2\pm14.0$&$-9.2\pm0.2$&$-2.4\pm1.6$&$7.9\pm1.0$&$0.0\pm0.2$&$-41.5\pm25.8$&$45.3\pm15.9$&$11.7\pm0.4$$\ $&$-0.8\pm1.6$&$1.3\pm1.0$&$0.9\pm0.3$&$62.5\pm26.9$&$1.6\pm1.7$\\
	C&$-926.6\pm13.4$&$551.1\pm7.9$&$-9.4\pm0.1$&$-0.7\pm1.5$&$6.8\pm0.9$&$1.5\pm0.1$&$-9.0\pm18.3$&$0.1\pm11.0$&$11.5\pm0.3$$\ $&$1.0\pm1.6$&$0.2\pm0.9$&$2.4\pm0.2$&$14.6\pm14.3$&$1.3\pm1.5$\\
	D&$-896.5\pm23.1$&$518.7\pm13.4$&$-19.6\pm0.5$&$-2.3\pm0.8$&$5.2\pm0.4$&$0.3\pm0.2$&$21.1\pm26.3$&$-32.2\pm15.4$&$1.3\pm0.6$$\ $&$-0.6\pm0.9$&$-1.4\pm0.5$&$1.2\pm0.3$&$38.6\pm22.3$&$0.9\pm0.9$\\
	E&$-882.9\pm28.6$&$518.5\pm16.8$&$-23.4\pm0.8$&$0.2\pm0.9$&$4.1\pm0.5$&$-1.5\pm0.3$&$34.7\pm31.2$&$-32.5\pm18.4$&$-2.5\pm0.8$$\ $&$1.9\pm1.1$&$-2.5\pm0.6$&$-0.6\pm0.4$&$47.6\pm29.2$&$3.1\pm1.5$\\
	F&$-866.7\pm57.5$&$518.8\pm34.4$&$-36.6\pm2.4$&$2.0\pm1.2$&$5.4\pm0.6$&$-2.5\pm0.8$&$50.9\pm58.8$&$-32.2\pm35.2$&$-15.7\pm2.4$$\ $&$3.7\pm1.3$&$-1.2\pm0.6$&$-1.6\pm0.8$&$62.2\pm48.5$&$4.0\pm2.3$\\
	Center & $-917.6 \pm 12.6$ & $550.9 \pm 7.5$& $-20.9 \pm 0.3$& $-1.7 \pm 0.5$& $6.6 \pm 0.3$& $-0.9 \pm 0.1$&\nodata&\nodata&\nodata&\nodata&\nodata&\nodata&\nodata&\nodata\\
	\enddata
	\tablecomments{Columns (2--7): 3D positions and motions of the subregions and reference center in the Heliocentric Galactic Cartesian coordinate system. The $x$-, $y$-, and $z$-axes are defined to point toward the Galactic center, the direction of Galactic rotation, and the north Galactic pole, respectively.
	Columns (8--13): 3D positions and motions of the subregions relative to the reference center. 
	Column (14): distance of the subregion to the reference center.
	Column (15): radial component of 3D velocities of the subregion relative to the reference center.
	}
\end{deluxetable*}

\subsection{Overall 3D Morphology and Motions}\label{sec:mor_motion}

\begin{figure*}[t!]
	\centering
	\includegraphics[width=0.9\textwidth]{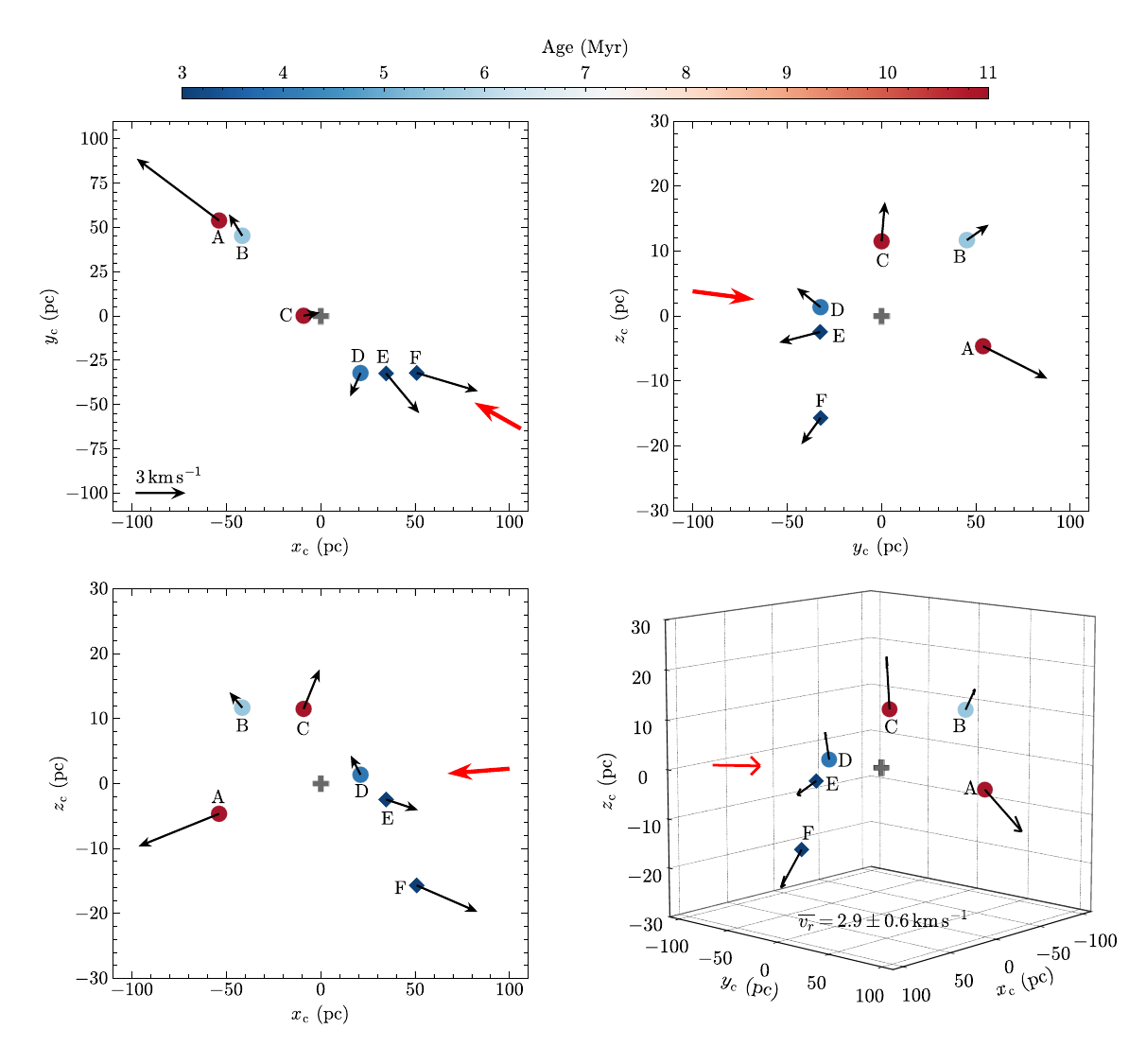}
	\caption{3D distributions and motions of the subregions in the local Cartesian coordinate system centered on the reference center of S205 (defined in Section~\ref{sec:6d_parameter_determination}), with its axes aligned with those of the Heliocentric Galactic Cartesian coordinate system. 
	The subregions are shown as circles and diamond symbols color-coded by ages, with relative motions shown as black arrows. The gray plus marks the reference center of S205. 
	Line-of-sight directions are indicated by red arrows. The upper-left, upper-right, lower-left, and lower-right panels show {the} $x_{\rm c}$-$y_{\rm c}$, $y_{\rm c}$-$z_{\rm c}$, $x_{\rm c}$-$z_{\rm c}$ projections, and {the} 3D view, respectively. The mean radial component of {the} 3D relative velocities is shown in the lower-right panel, with a positive value indicating expansion. The interactive version of {the} 3D view is available online. \label{fig:3d_project}}
\end{figure*}

%
To investigate the 3D structure and kinematics of S205, we derive the 
3D positions and velocities of {the} six subregions and the reference center
in the Heliocentric Galactic Cartesian coordinate system, 
as described in~Section~\ref{sec:6d_parameter_determination}.
The results are listed in Table~\ref{tab:3d_rel} (Columns 2--7).
The distance to the reference center, $1070\pm15$ pc, is consistent with 
the cloud distance of $\sim$1067$-$1114 pc measured by \citet{yan2021} 
using extinction and \gaia~DR2 parallaxes. 
Based on the 3D positions and velocities in the Heliocentric Galactic Cartesian 
coordinate system, we derive the corresponding 3D positions and velocities
in a local Cartesian coordinate system, 
following the procedure described in Section~\ref{sec:6d_parameter_determination}.
The results are listed in Table~\ref{tab:3d_rel} (Columns 8--13).

Figure~\ref{fig:3d_project} shows the 3D positions of the six subregions 
in the local Cartesian coordinate system. 
Unlike the CMa region, which is arranged approximately in a spherical shell \citep{dong2024}, 
the S205 subregions are not symmetrically distributed around the reference center, 
indicating that a single simple spherical or ellipsoid shell cannot describe its structure. 
The distances of the six subregions from the reference center, $\vert\boldsymbol{r}\vert$ 
(Column 14 of Table~\ref{tab:3d_rel}), span a wide range of $\sim$15--77 pc.

Figure~\ref{fig:3d_project} also shows the relative motions (black arrows) of the subregions. The velocity vectors exhibit systematic outward motions.
The radial components of the relative motions, $v_r$ (Column 15 of Table~\ref{tab:3d_rel}),
show that all six subregions are expanding, with a mean radial velocity of $2.9 \pm 0.6$~\kms. 
Following the method of \citet[Equation 5]{Grossschedl2026}, we further estimate the 
tangential components of {the} relative velocities $v_t$ of the subregions, 
with a mean value of $\sim$1.4 \kms. 
This value is smaller than that of the radial component, indicating that radial motions dominate the kinematics of the S205 region.
These results provide the first 3D evidence for the global expansion of S205. 
As indicated in Section~\ref{sec:6d_parameter_determination}, the mean RV of the molecular gas is assumed to represent the RV of each subregion, which may introduce relatively large uncertainties. Therefore, the derived 3D motions of S205 should be regarded as preliminary at this stage. More accurate determinations of the 3D motions of the subregions, particularly through additional high-precision stellar RV measurements, will be essential for further revising the 3D kinematics of the S205 region.

To test whether the global 3D expansion depends on the choice of reference center, 
we adopt subregions A and C as alternative reference centers, as they host the earliest-formed clusters in S205 (see cluster ages in Table~\ref{tab:basic_para}). 
We compare the values of $v_r$ and $v_t$ for the subregions relative to these 
reference centers, as shown in Figure~\ref{fig:vr_vt} of Appendix~\ref{app:test_AC}. 
The results consistently support the global expansion of S205. 
When subregion A is adopted as the reference center, $v_r$ exceeds $v_t$ for all 
other subregions. 
From this perspective, subregion A is a more likely candidate for the expansion center of S205.


\subsection{Internal Motions within Subregions}

\begin{figure*}[t!]
	\centering
	\includegraphics[width=1\textwidth]{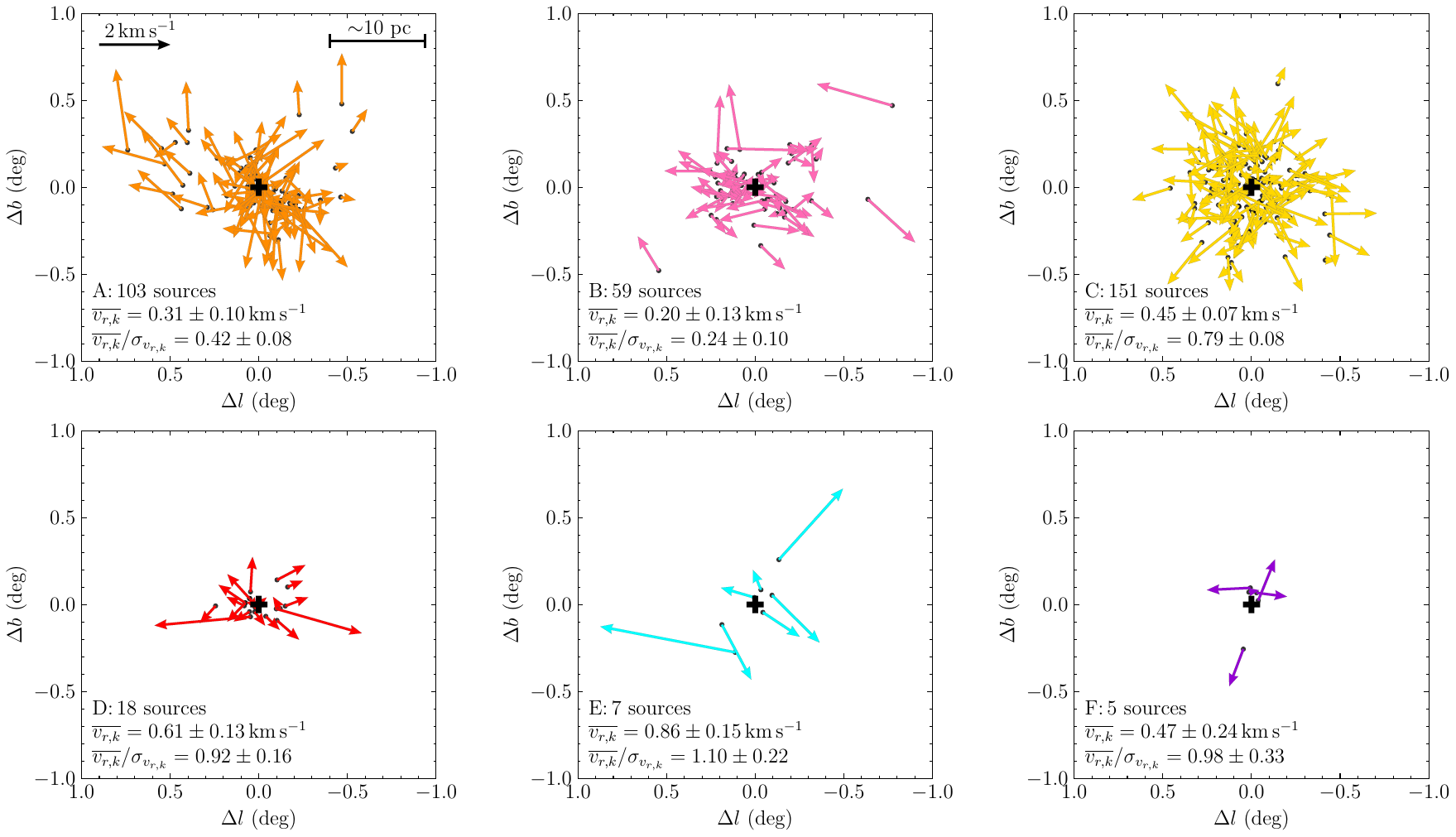}
	\caption{Internal motions of star clusters in each subregion. Black dots show the relative positions of young stars, and arrows show their residual tangential velocities after subtracting the mean motion of each cluster. Colors of subregions match Figure~\ref{fig:2dmotion} (b). Black crosses mark the cluster centers (Table~\ref{tab:basic_para}, Columns~3--4). The {lower}-left corner of each panel shows the number of young stars in this analysis, the mean radial component of the relative velocities $\overline{v_{r,k}}$, and its ratio to the corresponding dispersion $\overline{v_{r,k}} / \sigma_{v_{r,k}}$. \label{fig:subregion_expansion}}
\end{figure*}

The young stars in S205 {are} cluster members in six subregions. Previous studies indicate that young star clusters often exhibit coherent expansion in their early stages \citep[e.g.,][]{Kuhn2019,Della2024}. To test whether this feature is present in S205, we further analyze the internal kinematics of star clusters within each subregion, as described in Section~\ref{sec:method_3.2.2}.

Figure~\ref{fig:subregion_expansion} shows the residual tangential velocities of the cluster members after subtracting the overall motion of each cluster. Cluster members in all subregions exhibit systematic outward motions, with mean 2D expansion velocities ($\overline{v_{r,k}}$) of approximately 0.2--0.9 \kms, consistent with values found in similar young stellar systems \citep{Kuhn2019}. Thus, S205 exhibits both large-scale ($\sim$100 pc) global expansion (Section~\ref{sec:mor_motion}) and small-scale ($\sim$10 pc) expansion within subregions.

To assess the relative importance of ordered and disordered radial motions of the cluster members, we calculate the ratio of the mean radial component of the relative velocities to its corresponding dispersion, $\overline{v_{r,k}} / \sigma_{v_{r,k}}$. The resulting values (indicated in Figure~\ref{fig:subregion_expansion}) for the subregions are consistent with those of young clusters reported in \citet{Della2024}, suggesting that early expansion plays a significant role in the dynamical evolution of young stellar systems.

\section{Driving Sources of S205} \label{sec:sources}

\subsection{Momentum Estimation} \label{sec:momentum}

Massive young stars inject substantial momentum into the surrounding ISM through photoionization, stellar winds, and supernovae \citep[e.g.,][]{krumholz2014,Chevance2020}. 
To explore whether the global expansion of S205 is driven by such stellar feedback, we first estimate the momentum required to power the observed expansion, {as in,} e.g., \citet{orion2021} {and} \citet{posch2023}.
In Section~\ref{sec:mor_motion}, we derive outward radial velocities for the six subregions. We next compute the total masses of molecular gas and young stars.

Gas masses are estimated from the \uco~data. We adopt the commonly used CO-to-H$_2$ conversion factor
$X_{\rm CO}= 2.0 \times 10^{20}~\rm cm^{-2}(K\,km/s)^{-1}$  \citep{Bolatto_2013}
to convert {the} \uco~integrated intensities to {the} H$_2$ column density $N_{\rm H_2}$.
The total molecular mass is then
$M_{\rm gas}=2\mu m_{\scriptscriptstyle\rm H} a^2d^2\sum\nolimits_{i} N_{\rm H_2}$,
where $\mu=1.36$ is the mean atomic weight (considering helium and metals), $m_{\scriptscriptstyle\rm H}$ is the mass of a hydrogen atom, $a=30\arcsec$ is the angular size of {a pixel in} the \uco~data, $d$ is the distance in Column 6 of Table~\ref{tab:basic_para}, and the sum $\sum\nolimits_{i}$ runs over all MC pixels.
The resulting molecular masses are listed in Table~\ref{tab:basic_para} (Column~13).

To estimate the stellar mass, we do not simply sum the individual stellar masses inferred from CMD fitting (Section~\ref{sec:stellar_age_determination}). Instead, we construct the stellar mass function, modeled as $\frac{{\rm d}N_*}{{\rm d}M}\propto M^{-\alpha}$ \citep{Kroupa2001,Kroupa2024}. Assuming Poisson statistics for star counts $N_*$, the uncertainty in log($N_*$) is $1/(\sqrt{N_*}\ln 10)$, whose inverse is adopted as the weight in the fit. Here, binary effects are not considered. 
Following \citet{Hunt2024}, we integrate the mass function from 0.03~\msun~to the maximum stellar mass to yield the total stellar masses for subregions A--D (Table~\ref{tab:basic_para}, Column~14). These stellar masses are consistent with the values reported by \citet{Hunt2024} within uncertainties (Table~\ref{tab:ocs}). Subregions E and F contain only a few young stars, and their stellar {masses are} negligible compared to the molecular gas mass, contributing {little} to the momentum estimates.

Assuming that the radial momentum remains constant during the global expansion, the initial radial momentum can be inferred from the present-day value. Using the molecular gas and stellar masses of each subregion (Columns 13 and 14 of Table~\ref{tab:basic_para}) together with their radial expansion velocities relative to the center (Column 15 of Table~\ref{tab:3d_rel}), we obtain a total expansion momentum of $\sim$$1.1\pm0.3\times10^5$~\msun\,\kms~for the S205 region. 

Numerical simulations show that massive stars can inject momentum of up to $10^5$ \msun\,\kms~ into their natal cloud by photoionization and winds during {their} early evolution \citep[e.g.,][]{Haid2018}, while the {eventual} supernovae can add an additional 2--$4\times10^5$ \msun\,\kms~\citep[e.g.,][]{kim2015_SNe,walch2015}. Here, we assume that the feedback momentum is injected uniformly {into} a spherical shell and that only the cloud area exposed to the shock receives momentum. The current projected cloud area on the sky is $a_{\rm now}\sim1.3\times10^3\rm~pc^2$. Simplifying the cloud distribution into a spherical shell with surface area $A_{\rm shell}=4\pi \overline{\vert\boldsymbol{r}\vert}_{\rm now}^2$ and employing $\overline{\vert\boldsymbol{r}\vert}_{\rm now}\sim50\rm~pc$ from Table~\ref{tab:3d_rel}, the current exposed area fraction can be roughly estimated as $f_{\rm now}=a_{\rm now}/A_{\rm shell} \sim4\%$. Given an exposed area fraction of 2\%--8\% (i.e., $0.5$--$2\times f_{\rm now}$) at the onset of expansion, producing the momentum of the global expansion would require feedback from more than $\sim$10 massive stars or about three supernovae. Hence, identifying massive stars or their remnants in S205 is crucial to understanding the expansion mechanism, which is examined in the following section.

We note that this momentum analysis is a simplified estimate, and additional effects such as non-uniform motions, uncertain masses, unknown distances to the feedback source, unknown initial surface areas, or deviations from spherical symmetry could influence the derived values. For example, adopting a star formation efficiency of 10$\%$ \citep{Williams1997}, the total gas mass needed to produce the star clusters in S205 would be $\sim1.2\times10^4$~\msun, corresponding to approximately one-third of the current gas mass ($3.8\times10^4$~\msun). This suggests that the estimated momentum represents only a lower limit.

\subsection{Massive Stars}\label{sec:obstars}

We use the SIMBAD database\footnote{\url{https://simbad.u-strasbg.fr/simbad/}} \citep{simbad} to search for O--B1 type massive stars likely associated with the S205 region, including stars still within the region (\emph{Case 1}) and potential runaways (\emph{Case 2}).

\begin{figure}[t!]
	\centering
	\includegraphics[width=0.9\textwidth]{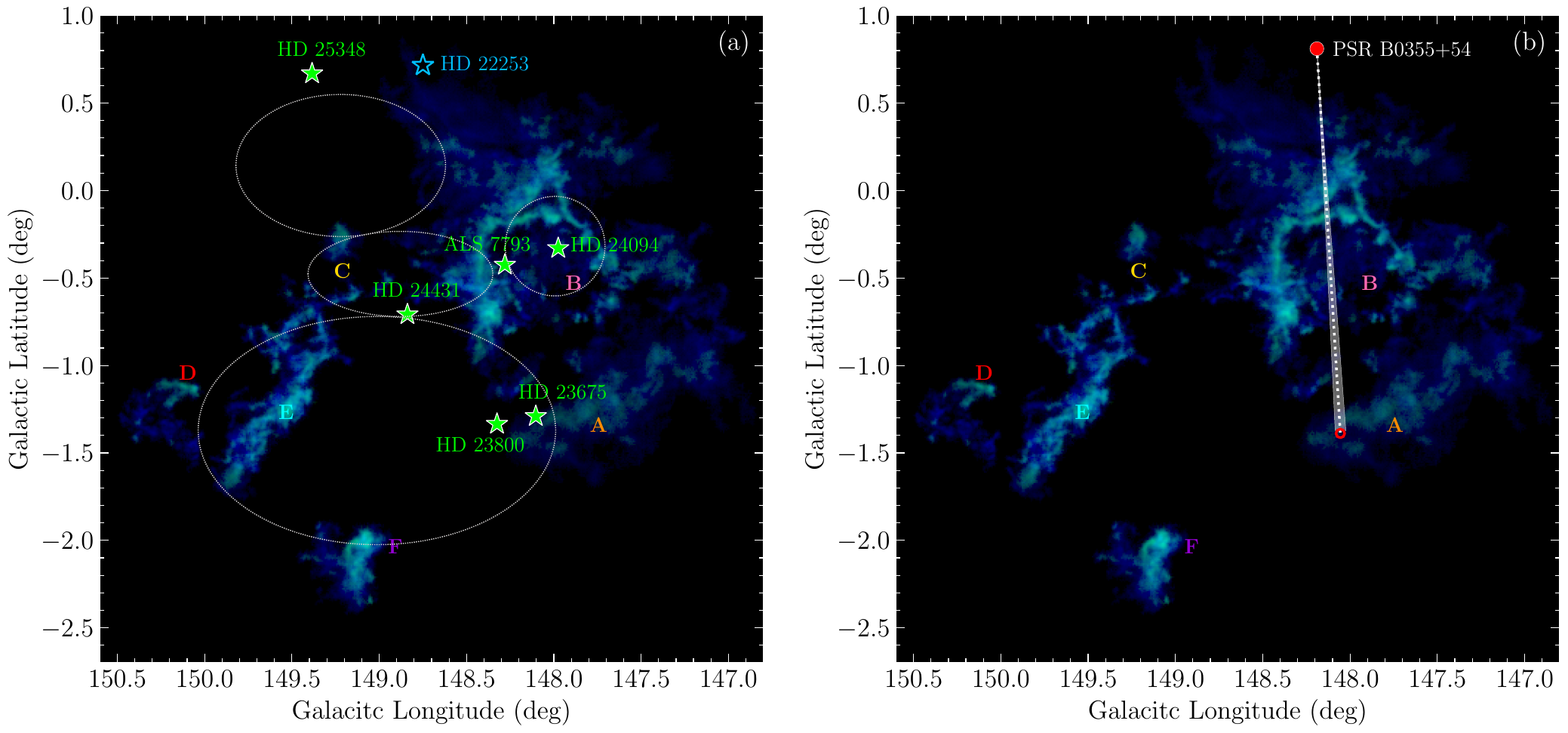}
	\caption{(a) O--B1 stars in S205. The green star symbols mark those within S205, and the blue star symbol is a trace-back target whose trajectory is shown in Appendix~\ref{app:auxiliary} Figure~\ref{fig:trace_OB} (b). The big circles are the same as those in Figure~\ref{fig:rgb}. (b) PSR B0355+54 and its trace-back trajectory (dashed line). Solid red circle: current position; open red circle: position 0.56 Myr ago (spin-down age of PSR B0355+54; \citealp{Taylor1975}). White shading indicates the trajectory uncertainty due to proper motion uncertainty. \label{fig:ob_pulsar}}
\end{figure}

\emph{Case 1}: Massive stars within the S205 region.
There are 12 massive stars within the $(l, b)$ boundaries of S205. {We set a relatively loose distance criterion of $<$500 pc from S205 to ensure the inclusion of all potential members. This criterion yields} six O--B1 stars: HD 24431 \citep[O9\,III$+$B1.5\,V;][]{maiz2021}, HD 23675 \citep[B0.5\,III;][]{HD23800}, HD 23800 \citep[B1\,IV;][]{HD23800}, ALS 7793 \citep[B1\,V;][]{als7793}, HD 25348 \citep[B1\,V;][]{HD25348}, and HD 24094 \citep[B1\,III, B8;][]{HD24094B8,HD24094B8B1III}.
Their spatial distributions are shown in Figure~\ref{fig:ob_pulsar} (a).

\emph{Case 2}: Potential runaway massive stars.
To hunt for stars possibly ejected from S205, we expand the search radius. Assuming a typical runaway velocity of $\sim$30 \kms~\citep[e.g.,][]{Blaauw1961_runaway1,Hoogerwerf2001}, corresponding to $\approx30~\rm pc\,Myr^{-1}$, stars could have moved $\sim$360 pc over 12 Myr (the S205 expansion timescale, see Section~\ref{subsection:timescale}). This tangential displacement is equivalent to an angular separation of $\sim$20\degree~at S205's distance. Adopting the S205 center ($149\fdg0, -1\fdg1$) as the search center and a search radius of 20\degree, we extract O--B1 stars from SIMBAD and cross-match {them} with \gaia~DR3, obtaining 421 stars. 
Restricting the sample to within 500 pc of S205 reduces it to 36 stars. 
To identify potential escapees, we trace the past trajectories of massive stars and thus require high-quality astrometry. After {excluding sources with} $\texttt{ruwe}>1.4$ {and} $\texttt{ipd\_gof\_harmonic\_amplitude}>0.1$, 35 massive stars remain. The angular separation of these stars from the S205 geometric center over time is shown in Appendix~\ref{app:auxiliary} Figure~\ref{fig:trace_OB} (a). We identify one candidate star, HD 22253 \citep[B0.5\,III;][]{HD23800}, that can be traced back to the S205 edge. Its trace-back position, trace-back time, and trajectory are shown in Figure~\ref{fig:trace_OB} (b). The \gaia~astrometric uncertainties are taken into account when calculating the trace-back trajectories. Because HD 22253 lacks RV measurements, we do not perform a full 3D trace-back analysis. The trace-back analysis assumes constant velocities for the massive stars, neglecting the effects of the Galactic gravitational potential and Galactic rotation. This approximation is generally valid for timescales of up to a few Myr and spatial scales of up to a few hundred pc \citep{Hoogerwerf2001}.

In summary, there are seven massive stars (Table~\ref{tab:obstar}) potentially formed in the S205 region, whose distributions are shown in Figure~\ref{fig:ob_pulsar} (a).
Among them, HD 24431, HD 23675, ALS 7793, and HD 24094 {have been} previously proposed as candidate ionizing sources for S205 \citep{Sharpless_1959,Avedisova1984,foster2006,Romero2008}, whereas this study adds HD 23800, HD 25348, and HD 22253 as new feedback candidates. In projection, HD 23675 and HD 23800 appear to be located in subregion A, and ALS 7793 and HD 24094 lie in subregion B. However, considering the distance uncertainties (Table~\ref{tab:obstar}), further data are required to confirm their true origin.

\begin{deluxetable*}{lcccccccccch}
	\tablecaption{Summary of Massive Star Candidates of S205 \label{tab:obstar}}
	\tablehead{\colhead{Name}&\colhead{\gaia~DR3}&\colhead{Spectral}	&\colhead{$ l $}&\colhead{$ b $}&\colhead{Distance} &\colhead{\pmra} &\colhead{\pmdec} &\colhead{\texttt{ruwe}}&\colhead{\texttt{ipd\_gof\_harmonic}}& \colhead{Mass}&\nocolhead{Ref.}\\
	\colhead{}&\colhead{\texttt{source\_id}}	&\colhead{Type}&\colhead{(\degree)}&\colhead{(\degree)}&\colhead{(pc)} &\colhead{\masyr}&\colhead{\masyr}&\colhead{}&\colhead{\texttt{\_amplitude}} &\colhead{(\msun)}&\nocolhead{}
	}
	\colnumbers
	\startdata
	HD 24431 & 251961924258399616 & O9\,III$+$B1.5\,V$^1$ & 148.84 & $-0.71$ & $964\pm54$ & $-1.77\pm0.05$ & $-0.05\pm0.05$ & $2.06$ & $0.17$ &  19.95$\pm$1.41$^2$ & 173 \\
	HD 23675 & 443789457348784640 & B0.5\,III$^3$ & 148.10 & $-1.29$ & $1079\pm26$ &$-0.39\pm0.02$ & $-1.88\pm0.02$ & $1.02$ & $0.03$ &  19.33$\pm$0.25$^2$ & 92 \\
	HD 23800 & 251618704831632256 & B1\,IV$^3$ & 148.33 & $-1.33$ & $659\pm28$ & $-1.63\pm0.07$ & $-2.22\pm0.06$ & $2.90$ & $0.02$ & 10.00$\pm$0.58$^2$  & 81 \\
	ALS 7793 & 252046998970045184 & B1\,V$^4$ & 148.28 & $-0.43$ & $1209\pm75$ & $-0.42\pm0.06$ & $-2.18\pm0.04$ & $4.07$ & $0.03$ &  11.98$^2$, 11.8$^5$  & 19 \\
	HD 25348 & 275918251105986176 & B1\,V$^6$ & 149.38 & $0.67$ & $1051\pm25$ & $0.02\pm0.02$ & $-1.11\pm0.02$ & $0.89$ & $0.02$ &  11.98$^2$, 11.8$^5$  & 62 \\
	HD 24094 & 444210776459268736 & B1\,III$^{7}$, B8$^{7,8}$ & 147.97 & $-0.33$ & $1160\pm58$ & $-1.00\pm0.05$ & $-1.51\pm0.04$ & $1.84$ & $0.02$ &  11.98$^2$ &   \\
	HD 22253 & 448769351668458368 & B0.5\,III$^3$ & 144.28 & $0.92$ & $703\pm15$ & $-2.26\pm0.04$ & $-0.29\pm0.03$ & $1.07$ & $0.01$ &  27.85$\pm$3.07$^2$ & 77 \\
	\enddata
	\tablecomments{References: 1. \citet{maiz2021}, 2. \citet{Hohle2010}, 3. \citet{HD23800}, 4. \citet{als7793}, 5. \citet{Pecaut2013}, 6. \citet{HD25348}, 7. \citet{HD24094B8B1III}, and 8. \citet{HD24094B8}. For HD 24431, HD 23675, HD 23800, and HD 22253, {the} masses are taken from {the catalog compiled by} \citet{Hohle2010}. For ALS 7793, HD 25348, and HD 24094, {the} masses {are adopted from the} typical values {corresponding to} their {respective} spectral types \citep{Hohle2010,Pecaut2013}.}
\end{deluxetable*}

\subsection{Remnants of Massive Stars: Pulsars}\label{sec:pulsars}

In addition to the existing massive stars, remnants {of} their terminal evolution, e.g., pulsars, can provide clues to the feedback history. Pulsars are the products of massive star supernovae \citep[e.g.,][]{Scheck2006}, and their characteristic ages (spin-down ages) can be estimated from rotational period models. We search the S205 region and its surroundings using the ATNF Pulsar Catalogue\footnote{\url{https://www.atnf.csiro.au/research/pulsar/psrcat}} \citep[v2.6.5;][]{ATNFpulsar}. Typical tangential velocities of pulsars are $\sim$300 \kms~\citep[e.g.,][]{Lyne1994,Hobbs2005}, giving a tangential displacement of up to $\sim$3600 pc over 12 Myr (the S205 expansion timescale, see Section~\ref{subsection:timescale}), corresponding to $\sim$70\degree~on the sky. Candidates are selected with spin-down ages $\leq$12 Myr, available astrometry, and positions within 70\degree~of the S205 center $(l=149\fdg0, b=-1\fdg1)$, yielding 29 pulsars. {The} Galactic potential has minimal effect on motions over {timescales of} $\leq$10 Myr \citep{Noutsos2013}. Therefore, we project proper motion vectors onto the $l$-$b$ plane and visually inspect their directions to exclude sources not ejected from S205, leaving six candidates.

Following the method of \citet{Bobylev2009}, we trace pulsar motions backward using their spin-down ages. The trajectory of PSR B0355+54 \citep{Taylor1972,Davies1972,Manchester1972} can be traced back to subregion A of S205, as shown in Figure~\ref{fig:ob_pulsar} (b), suggesting a possible physical association. Its spin-down age is 0.56 Myr \citep{Taylor1975}. Pulsar timing measurements from \citet{Li2016_pulsar_pm} give a proper motion of $\mu_{\alpha^{*}}=9.3\pm0.5$ \masyr~and $\mu_{\delta}=8.3\pm0.8$ \masyr, corresponding to a tangential velocity of $\sim$73 \kms~relative to the S205 center. VLBA observations \citep{Chatterjee2004} yield a parallax distance of $1.04^{+0.21}_{-0.16}$ kpc, consistent with S205, and FAST scintillation arc measurements \citep{Ocker2024_distance} indicate that scattering screen distances may also coincide with S205. These results strongly suggest that PSR B0355+54 could be the remnant of a massive star formed in S205.

To conclude, S205 is found to harbor seven candidate massive stars (HD 24431, HD 23675, ALS 7793, HD 24094, HD 23800, HD 25348, HD 22253) and one potential remnant of a massive star, the pulsar PSR B0355+54. As shown in Section~\ref{sec:momentum}, the combined feedback from these objects may account for the overall expansion of S205. 
In addition, the cometary-shaped cloud structures in subregions D and F (Figure~\ref{fig:ob_pulsar}), together with the outward motions of their associated young stars (Figure~\ref{fig:2dmotion}), suggest feedback-influenced morphology and star formation. Such phenomena have been reported in the Orion region \citep{orion2021} and the Corona Australis (CrA) region \citep{posch2023}.

\section{Star Formation History in S205}
\label{section:history}

\subsection{The Dawn of Star Formation} \label{subsection:timescale}

\begin{figure*}[t!]
	\centering
	\includegraphics[width=0.9\textwidth]{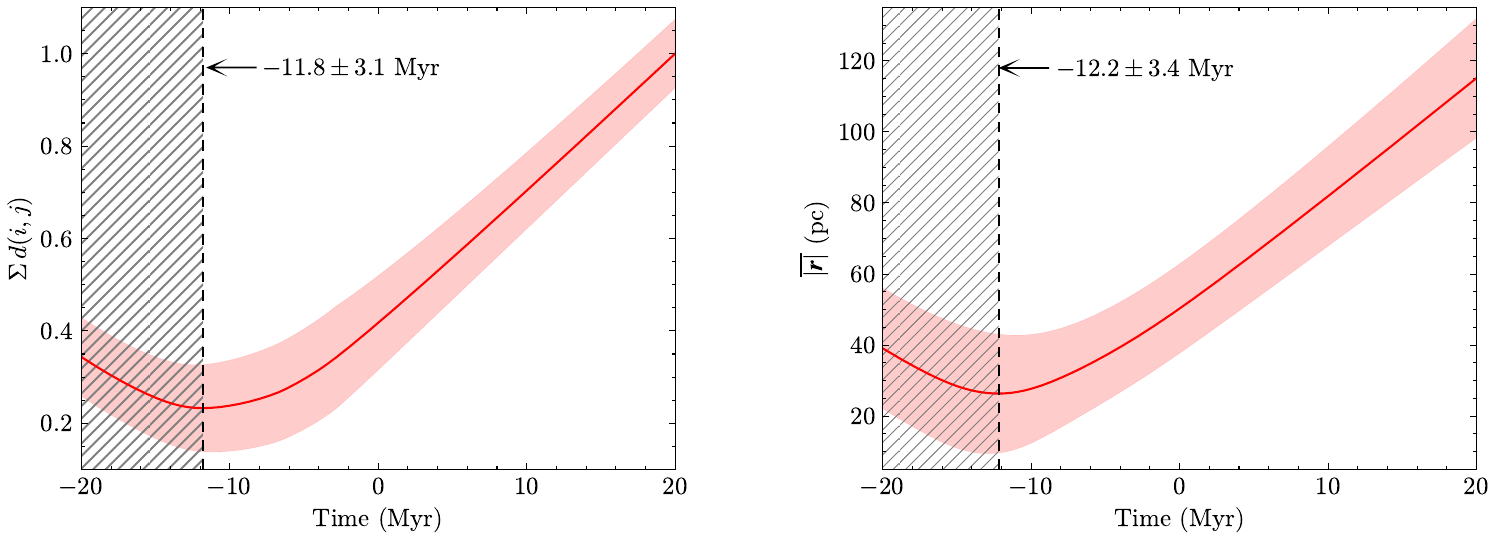}
	\caption{Left: Evolution of the sum of distances between the six subregions, $\Sigma d{(i,j)}$, normalized by its maximum. Right: Evolution of the mean distance of the six subregions from the S205 reference center, $\overline{\vert\boldsymbol{r}\vert}$. Time range is $\pm20$ Myr. Vertical dashed lines in two panels indicate the times when $\Sigma d{(i,j)}$ and $\overline{\vert\boldsymbol{r}\vert}$ reach minimum values, respectively. \label{fig:traceback}}
\end{figure*}

%
In this section, we explore the star formation history of S205. To trace the origin of star formation, we perform a trace-back analysis of the motions of its subregions. Despite the complex, bubble-dominated {morphology}, combining the data of young stars and MCs reveals a systematic expansion of S205. We therefore follow the approach of \citet{orion2021}, integrating the orbits of the subregions backward and forward in time (from $-$20 Myr to $+$20 Myr) to identify the moment of their most compact configuration. This moment can be interpreted as the onset of star formation and stellar feedback in S205.

Assuming the subregions move at constant velocities, we estimate the onset of global expansion using two independent methods to test the robustness of the results. Following \citet{orion2021}, the first method computes the sum of distances in 3D Cartesian space between {the} six subregions, $\Sigma d{(i,j)}$, and identifies the time of its minimum as the start of expansion. The second method tracks the mean distance of the six subregions from the S205 reference center, $\overline{\vert\boldsymbol{r}\vert}$, and takes {the time of} its minimum as the expansion onset. Both methods use a time step of 0.1 Myr. 

The evolutionary {tracks} are shown in Figure~\ref{fig:traceback}, and the subregion positions during traceback in the projections ($x_{\rm c}$-$y_{\rm c}$, $x_{\rm c}$-$z_{\rm c}$, $y_{\rm c}$-$z_{\rm c}$) are shown in Figure~\ref{fig:trajectory} of Appendix~\ref{app:auxiliary}. The two methods yield consistent results, $-11.8\pm3.1$~Myr and $-12.2\pm3.4$~Myr. At that time, the mean distance of {the} six subregions from the reference center is $\sim${27} pc, about half the current value ($\sim$50 pc). This indicates that star formation in S205 began at least 11--12 Myr ago.

\subsection{Multiple Episodes of Star Formation}\label{sec:ages}

\begin{figure}[t!]
	\centering
	\includegraphics[width=0.9\textwidth]{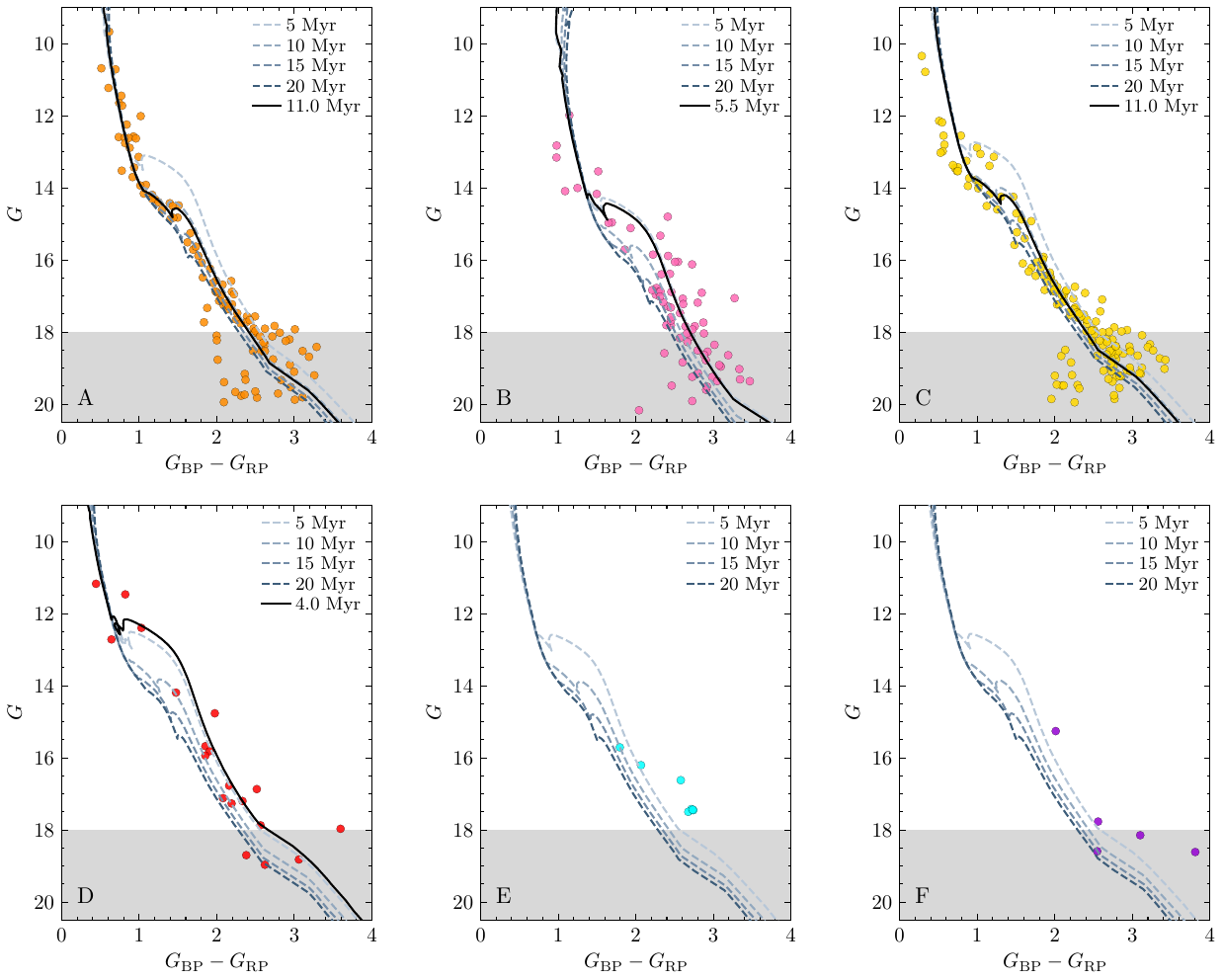} 
	\caption{CMDs of young stars in each subregion. Black curves show the best-fit isochrones for subregions A--D. {Stars} fainter than 18 mag (gray shading area) are excluded from the CMD fitting. Subregions E and F are not fitted because of the limited number of young stars. The blue-gray dashed curves indicate the reference theoretical isochrones. \label{fig:regions}}
\end{figure}

Using the method described in Section~\ref{sec:stellar_age_determination}, the ages of subregions A--D are derived from CMD fitting (Figure~\ref{fig:regions}). Column (15) of Table~\ref{tab:basic_para} presents the fitted cluster ages. Although our fitted ages differ slightly from those reported by \citet{Hunt2024yCat}, they remain consistent within the uncertainties. The differences may arise from the inclusion of additional YSOs in our membership sample and from fitting only stars brighter than $G = 18$ mag. Subregions E and F contain too few stars to allow reliable age estimates. We therefore adopt an indicative age of $\sim$3 Myr, considering the typical ages of Class II YSOs \citep[e.g.,][]{Williams2011}. In Figure~\ref{fig:regions}, we show the reference theoretical isochrones for subregions E and F. The distance moduli are derived from their parallax-based distances, and the extinction and reddening parameters are assumed to be the mean values of subregions C and D, which are spatially closest to E and F. Although subregions E and F contain only a small number of young stars, their assumed $\sim$3 Myr ages appear to be reasonable.

Subregions A and C have ages of $\sim$11 Myr, in agreement with the onset of star formation inferred from the trace-back analysis. This result not only validates the trace-back method, but also reinforces that star formation in S205 commenced 11--12 Myr ago. Compared to A and C, subregions B and D are younger, with ages of 4--5 Myr, indicating a second episode of star formation. 
In particular, the ages of young stars in subregion B match the expansion age of bubble LBN 148.11$-$0.45 \citep[$\sim$4 Myr;][]{Romero2008}, supporting a ``collect-and-collapse'' triggering scenario \citep{Romero2009}.
Overall, S205 has experienced at least two episodes of star formation.

The search for massive stars in Section~\ref{sec:sources} also {supports} the multi-{episode} star formation scenario. First, O-type stars typically no longer exist after 5--6 Myr \citep[e.g.,][]{Meynet2003,Weidner2010,Berlanas2025}, so the O9 star HD 24431 likely represents the second episode of star {formation that occurred} $\sim$4--5 Myr ago. Second, B0.5--B1 stars (HD 23675, ALS 7793, HD 24094, HD 23800, HD 25348, 
HD 22253) have ages $\lesssim$8--20 Myr \citep[e.g.,][]{Levenhagen2004,Daszynska-Daszkiewic2019,Liu2023}, broadly consistent with the first episode of star formation activity $\sim$11--12~Myr ago. 
While these stars likely originated during this earlier epoch, a more recent origin during the subsequent stage cannot be entirely ruled out.

We infer the evolutionary scenario for S205 as follows: an early episode of massive stars and star clusters formed about 11--12 Myr ago, likely occurring in two star-forming subregions (subregions A and C), and drove the global expansion, followed by a subsequent star-forming episode $\sim$4--5 Myr ago; a supernova explosion event $\sim$0.56~Myr ago likely further accelerated the region's evolution. These processes collectively produce recurrent star formation activity.

\section{Summary}\label{sec:summary}

Integrating high-precision astrometry of YSOs and young star cluster members from \gaia~DR3 with high-quality \uco~and \lco~(1--0) molecular line data from the MWISP survey, we present the first systematic study of the 3D structure, kinematics, and evolutionary history of the star-forming regions in the vicinity of the \ion{H}{2} region S205, and provide new constraints on its star formation processes. The main conclusions are as follows:

\begin{enumerate}
\item Based on \gaia~DR3, we find that S205 lies at an average distance of about $1070\pm15$ pc. Young stars in S205 closely trace the MCs, illustrating a complex 3D structure composed of multiple substructures at distances between about 1000 {and} 1150 pc. Kinematic analysis further shows that S205 exhibits expansion at two scales: {the} global expansion of the entire region and the local expansion of substructures.

\item Based on the mass and expansion velocity, the radial momentum of S205 is estimated to be $\sim$$1.1\times10^5$~\msun\,\kms. We identify seven massive stars (O9--B1) potentially formed in S205 and one supernova remnant candidate, the pulsar PSR B0355+54. The combined feedback from these sources may account for the observed global expansion of the region.

\item Trace-back analysis {suggests that} star formation in S205 likely started about 11--12 Myr ago, consistent with the $\sim$11 Myr age derived from CMD fitting {for} two star clusters. Two additional star clusters have ages of $\sim$4--5 Myr, indicating that S205 has experienced at least two episodes of star formation.

\end{enumerate}

To conclude, S205 shows a plausible multi-episode star formation scenario: an early episode of star formation drove the global expansion, followed by subsequent star formation triggered by local stellar feedback, with a more recent supernova explosion further accelerating the region's evolution.

\begin{acknowledgments}
We thank the referee for the constructive suggestions that helped to improve this paper. This work was funded by the National Key R\&D Program of China (grant No.\,2024YFA1611504), the Xinjiang Talent Development Fund (grant No.\,XJRC-2025-KJ-YJ-CXPT-180), and the National SKA Program of China (grant No.\,2022SKA0120103). C. H. acknowledges support from the National Postdoctoral Program for Innovative Talents of the Office of China Postdoc Council (grant No.\,BX20240414) and the NSFC grant No.\,12403041. Y. L. thanks the support of the NSFC grant No.\,12203104. Z. L. thanks the support of the NSFC grant No.\,12403077. D. L. thanks the support of the NSFC grant No.\,12503071.
This research used the data from the Milky Way Imaging Scroll Painting (MWISP) project, which is a multiline survey in \uco/\lco/\ltco~along the northern Galactic plane with the PMO 13.7 m telescope. MWISP was sponsored by the National Key R\&D Program of China with grant 2017YFA0402701 and the CAS Key Research Program of Frontier Sciences with grant QYZDJ-SSW-SLH047. 
This work has made use of data from the European Space Agency (ESA) mission
{\it Gaia} (\url{https://www.cosmos.esa.int/gaia}), processed by the {\it Gaia}
Data Processing and Analysis Consortium (DPAC,
\url{https://www.cosmos.esa.int/web/gaia/dpac/consortium}). Funding for the DPAC
has been provided by national institutions, in particular the institutions
participating in the {\it Gaia} Multilateral Agreement.
This research has made use of the VizieR, SIMBAD, and Aladin databases operated at CDS, Strasbourg, France.
\end{acknowledgments}

\software{Astropy \citep{Astropy2013,Astropy2018}, matplotlib \citep{Hunter_2007}, NumPy \citep{Numpy}, SciPy \citep{SciPy}, pandas \citep{pandas}, TOPCAT \citep{TOPCAT}}, Plotly \citep{plotly}.

\appendix
\section{Auxiliary Figures} \label{app:auxiliary}
The auxiliary figures are presented in the order of their appearance in the main text.

\begin{figure}[ht!]
	\centering
	\figurenum{A1}
	\includegraphics[width=0.4\textwidth]{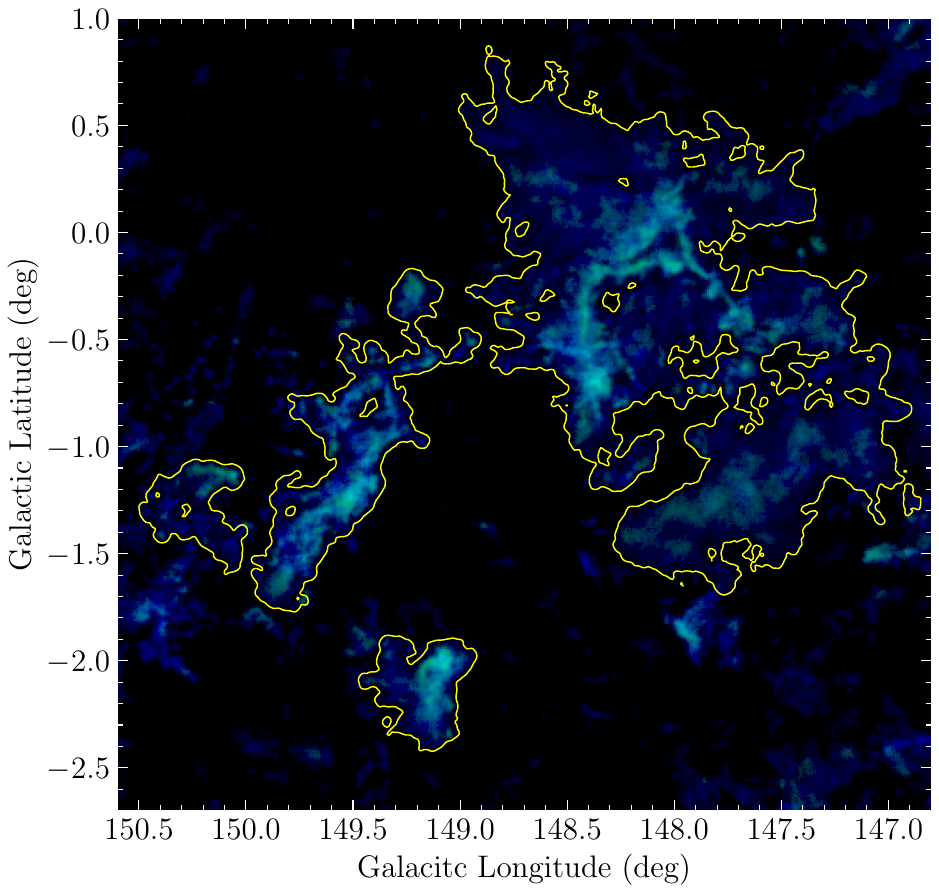}
	\caption{Molecular gas distribution toward S205. Blue and green represent the integrated intensity maps of \uco~and \lco~emission, respectively. Yellow contours highlight \uco~boundaries of five clouds listed in Table~\ref{tab:mcs}. See Section~\ref{sec:cloud}. \label{fig:allv}}
\end{figure}

\begin{figure*}[ht!]
	\centering
	\figurenum{A2}
	\includegraphics[width=0.65\textwidth]{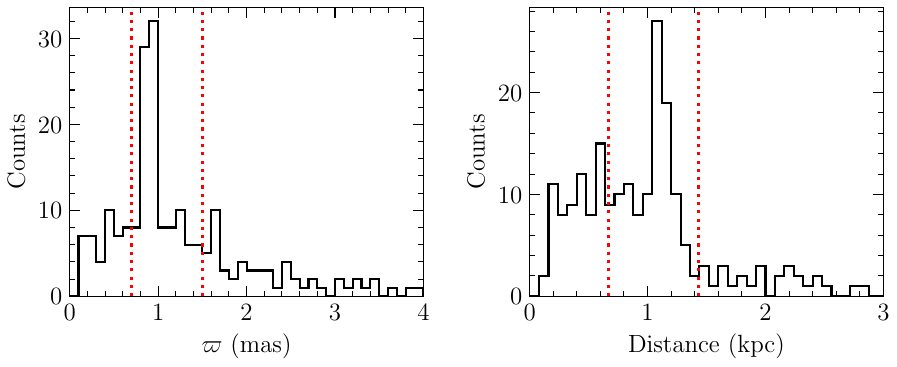}
	\caption{Distributions of parallaxes (left) and derived distance (right) for the initial YSO sample as collected from the literature (see Section~\ref{sec:ysos}). The adopted parallax cut and its corresponding distance limits are indicated by red dashed lines. \label{fig:plx}}
\end{figure*}

\begin{figure*}[ht!]
	\centering
	\figurenum{A3}
	\includegraphics[width=1\textwidth]{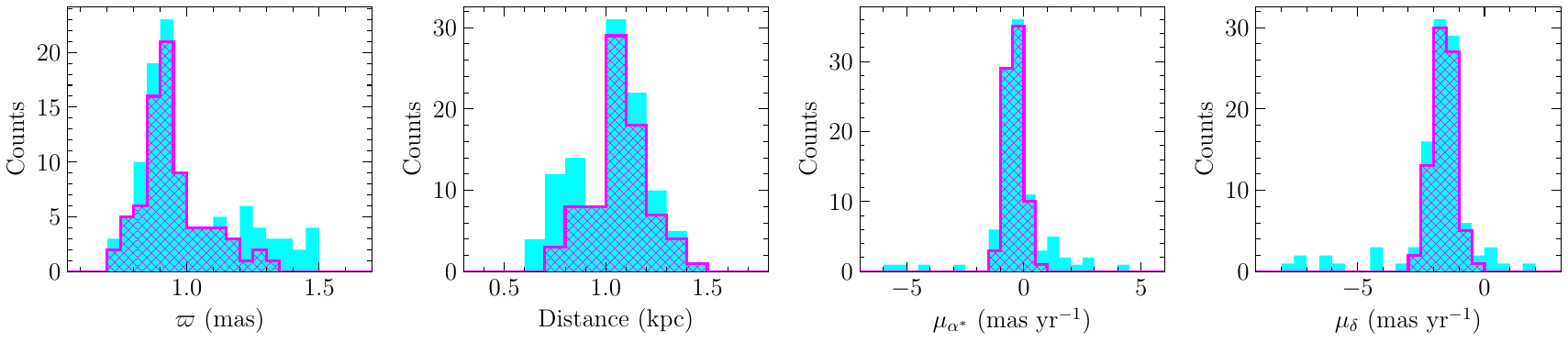}
	\caption{From left to right: distributions of parallax, distance, and proper motions (\pmra~and \pmdec) of the selected YSOs (cyan) and after outlier rejection (magenta). See Section~\ref{sec:ysos}. \label{fig:plx_pm}}
\end{figure*}

\begin{figure*}[ht!]
	\centering
	\figurenum{A4}
	\includegraphics[width=1\textwidth]{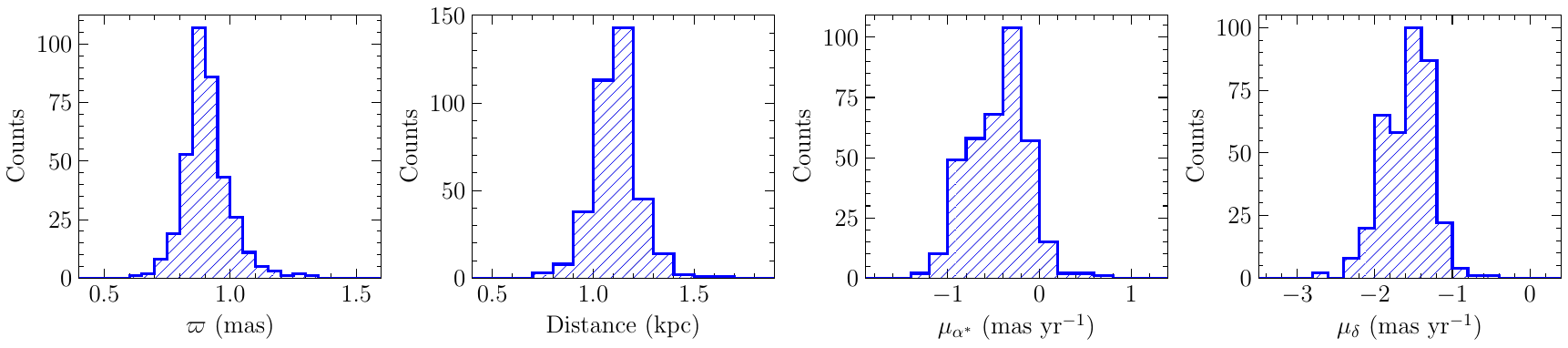}
	\caption{Same parameters as in Figure~\ref{fig:plx_pm}, but for the full young star sample, including both YSOs and young cluster members (see Section~\ref{sec:ocs}). \label{fig:a4}}
\end{figure*}

	\begin{figure}[t!]
	\centering
	\figurenum{A5}
	\includegraphics[width=0.8\textwidth]{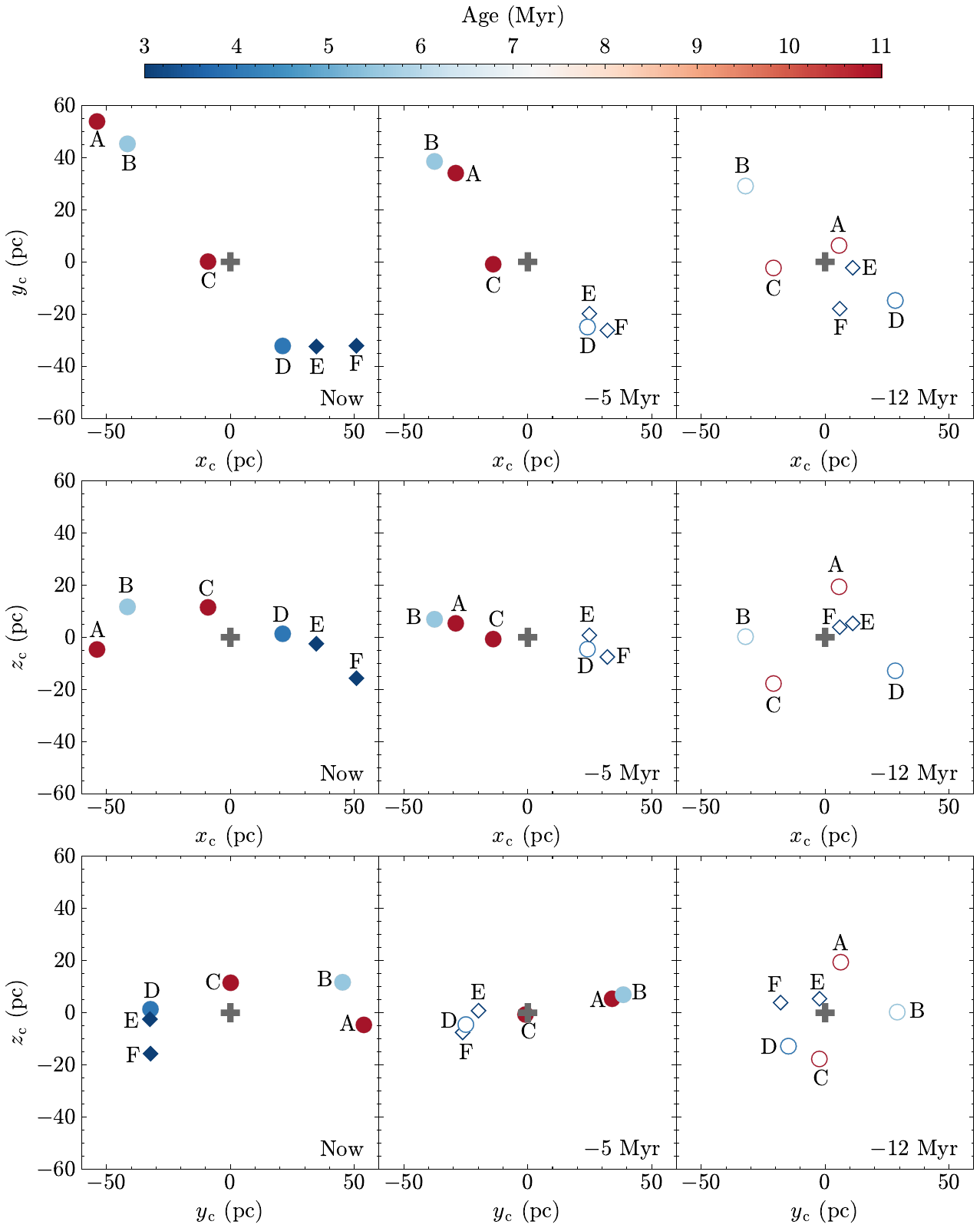}
	\caption{3D distribution of subregions at the present and at two times in the past ($-$5 Myr and $-$12 Myr), representing the approximate times of the two star formation episodes (see Section~\ref{sec:ages}). The time is labeled in the bottom-right corner of each panel. 
	From top to bottom, we show projections onto the $x_{\rm c}-y_{\rm c}$, $x_{\rm c}-z_{\rm c}$, and $y_{\rm c}-z_{\rm c}$ planes. The symbols and colors are as in Figure~\ref{fig:3d_project}. Open symbols indicate subregions whose stars had not yet formed by the corresponding trace-back times. The interactive 3D version of this figure is available online.
	\label{fig:trajectory}}
	\end{figure}

\begin{figure}[ht!]
	\centering
	\figurenum{A6}
	\hspace{0.8cm}\includegraphics[width=0.9\textwidth]{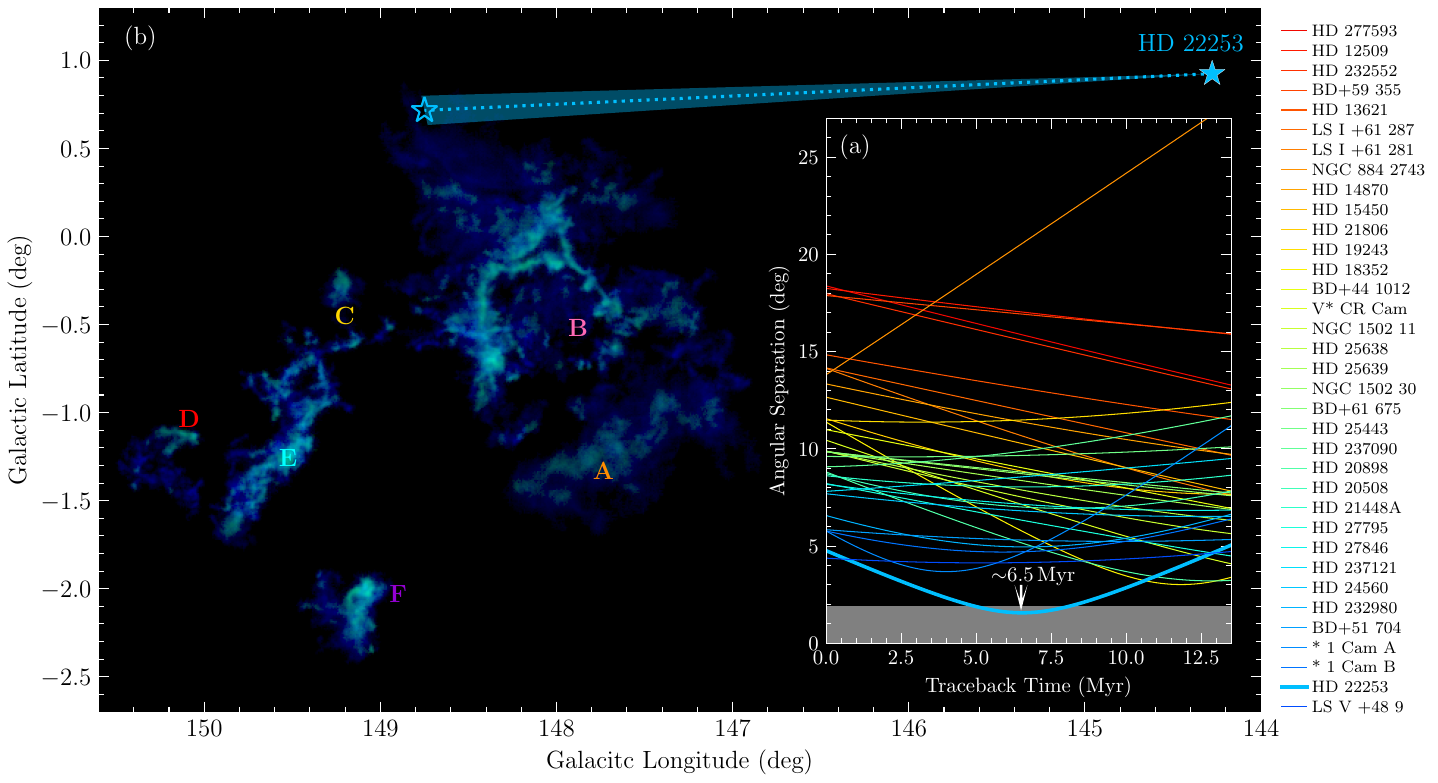}
	\caption{(a) Angular distance of OB stars around S205 {from} its geometric center as a function of trace-back time. The gray shaded area corresponds to locations inside the S205 boundary. The trace-back curve of HD 22253 is highlighted, with the time of closest approach (6.5 Myr ago) indicated. (b) Trace-back trajectory of HD 22253 relative to the S205 center (blue dashed line), with the hollow star symbol marking its position 6.5 Myr ago. See Section~\ref{sec:obstars}. \label{fig:trace_OB}}
\end{figure}

\section{Comparison Between Stellar and Molecular Gas Radial Velocities} \label{app:rvs}

RV measurements of the young stars compiled in Section~\ref{sec:young_stars} are collected from \gaia~DR3 and supplemented by cross-matching with the SDSS-V DR19 products, including the Milky Way Mapper \citep[MWM,][]{MWM2025} and BOSS catalogs \citep{SDSS-V}. Only RV measurements with associated uncertainties are retained. For young stars with RVs available from multiple catalogs, we adopt the measurement with the smallest uncertainty. This yields RV measurements for 59 young stars, comprising 16\% of the young star sample. To facilitate comparison with the molecular gas velocities, the heliocentric stellar RVs are converted to the conventional LSR frame, which assumes a solar motion of 20 \kms~toward $\alpha=18^{\rm_h}$, $\delta=+30\arcdeg$ \citep[B1900;][]{Gordon1976}.

\begin{figure}[ht!]
	\centering
	\figurenum{B1}
	\includegraphics[width=1\textwidth]{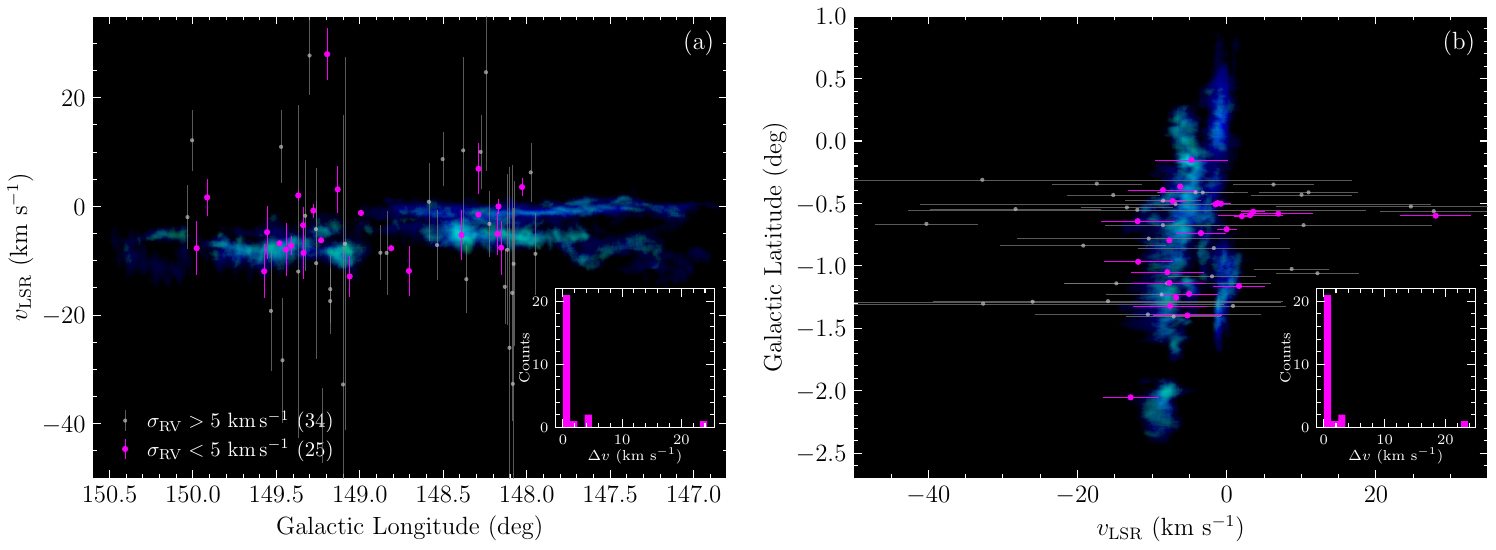}
	\caption{PV diagrams for $l$--$v$ (a) and $b$--$v$ (b) of the \vlsr~of the molecular gas (\uco: blue, \lco: green) and young stars in S205. Gray and magenta circles denote young stars with RV uncertainties $\sigma_{\rm RV}>5$ \kms~and $<5$ \kms, respectively, with the numbers in parentheses indicating the corresponding sample sizes. 
	The inset panels show the distributions of star-gas velocity separation $\Delta v$ from the $l$--$v$ and $b$--$v$ maps, respectively. See also Section~\ref{sec:6d_parameter_determination}.
	 \label{fig:lv_bv}}
\end{figure}

Figure~\ref{fig:lv_bv} presents the position-velocity (PV) diagrams for $l$--$v$ and $b$--$v$ of the young stars with RVs and the \vlsr~of the molecular gas from the whole studied region. 
The RV measurements of most young stars have large uncertainties. Twenty-five stars (comprising 7\% of the young star sample) have relatively small RV uncertainties ($\sigma_{\rm RV}<5$ km s$^{-1}$). For each of these stars, we calculate $\Delta v=\rm |RV_{star}-RV_{gas}|$, i.e., the RV separation between the young star (accounting for stellar RV uncertainty) and its nearest molecular gas on the $l$--$v$ or $b$--$v$ map. If the RV of a star falls within the RV range of the gas, $\Delta v$ is defined to be zero. As shown in the inset panels of Figure~\ref{fig:lv_bv}, $\Delta v$ values of most young stars are close to zero. Overall, 22 of the 25 young stars (88\%) have $\Delta v<2$ \kms, comparable to the RV offsets between YSOs and molecular clouds reported by \citet{YangL2025}. These results indicate that the RVs of young stars in S205 are broadly consistent with those of molecular clouds within the uncertainties.

We compile the RVs of the young stars and molecular gas in each subregion, as listed in Table~\ref{tab:subregion_rvs}. The subregions are defined in Section~\ref{sec:6d_subregion}. The gas RVs of the subregions are derived from the \lco~emission and are the same as those listed in Column (11) of Table~\ref{tab:basic_para}. The stellar RV of each subregion is defined as the mean RV of the young stars within that subregion, and the uncertainty is estimated using a Monte Carlo sampling method that accounts for the RV measurement uncertainties. To evaluate the impact of the stellar RV uncertainties, we consider three cases: no threshold, $\sigma_{\rm RV}<5$ \kms, and $\sigma_{\rm RV}<3$ \kms.

\begin{figure}[ht!]
	\centering
	\figurenum{B2}
	\includegraphics[width=0.75\textwidth]{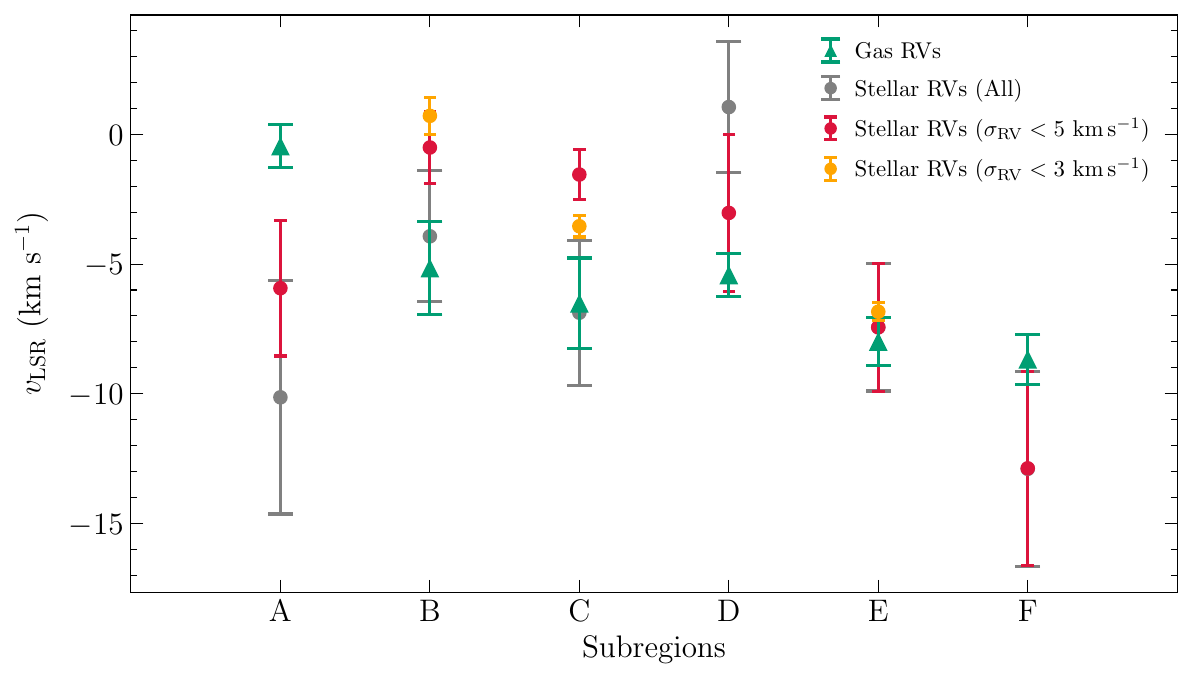}
	\caption{Comparison between the stellar and gas RVs in each subregion. The gas RVs of the subregions are derived from the \lco~emission. The stellar RV of each subregion is calculated as the mean RV of the young stars within that subregion. Three cases are included: no threshold, $\sigma_{RV}<5$ \kms, and $\sigma_{RV}<3$ \kms. The corresponding values are listed in Table~\ref{tab:subregion_rvs}.
	 \label{fig:subregion_rvs}}
\end{figure}

The comparison between the stellar and gas RVs in each subregion is shown in Figure~\ref{fig:subregion_rvs}. The current data only tentatively {suggest} that the stellar and gas RVs in the subregions lie within similar ranges. In some subregions (e.g., A and C), different stellar RV subsamples appear to deviate from the gas RVs or show sensitivity to the adopted uncertainty thresholds. These differences may arise from several factors. First, the limited number of stellar RV measurements and possible systematic differences among RVs from different surveys make it difficult to obtain robust mean stellar RVs. Second, most stellar RV measurements have large uncertainties, leading to substantial uncertainties in the derived mean stellar RVs. Third, there may be real velocity deviations between the gas and young stars in these subregions. Therefore, further investigations, especially those based on {additional} stellar RV measurements, are required to better assess the close kinematic connection between young stars and molecular gas in S205.

\begin{deluxetable*}{chcccccccccc}
	\setlength{\tabcolsep}{0.8mm}
	\tablenum{B1}
	\tabletypesize{\footnotesize}
	\tablecaption{RVs of Young Stars and Molecular Gas in Subregions \label{tab:subregion_rvs}}
	\tablehead{\colhead{Subregion}&\nocolhead{$N_*$}&\colhead{$N_{Gaia}$}	&\colhead{$N_{\rm MWM}$}&\colhead{$N_{\rm BOSS}$}&\colhead{$N\rm (All)$} &\colhead{RV(All)} &\colhead{$N({\sigma_{\rm RV}<5})$} &\colhead{RV$({\sigma_{\rm RV}<5})$}&\colhead{$N({\sigma_{\rm RV}<3})$} &\colhead{RV$({\sigma_{\rm RV}<3})$}&\colhead{RV(gas)}\\
	\colhead{}&\nocolhead{}	&\colhead{}&\colhead{}&\colhead{}&\colhead{} &\colhead{(\kms)}&\colhead{}&\colhead{(\kms)}&\colhead{} &\colhead{(\kms)}&\colhead{(\kms)}
	}
	\colnumbers
	\startdata
	A&107&7&5&7&13&$-10.1\pm4.5$&3&$-5.9\pm2.6$&0&\nodata&$-0.5\pm0.8$\\
	B&72&4&8&4&12&$-3.9\pm2.5$&5&$-0.5\pm1.4$&3&$0.7\pm0.7$&$-5.2\pm1.8$\\
	C&158&15&19&7&27&$-6.9\pm2.8$&12&$-1.6\pm0.9$&6&$-3.5\pm0.4$&$-6.5\pm1.7$\\
	D&19&3&1&1&4&$1.1\pm2.5$&2&$-3.0\pm3.0$&0&\nodata&$-5.4\pm0.8$\\
	E&7&0&1&1&2&$-7.5\pm2.5$&2&$-7.5\pm2.5$&1&$-6.8\pm0.3$&$-8.0\pm0.9$\\
	F&5&1&0&1&1&$-12.9\pm3.8$&1&$-12.9\pm3.8$&0&\nodata&$-8.7\pm1.0$\\
	\enddata
	\tablecomments{Columns (2)--(4): numbers of young stars with RV measurements from \gaia, MWM, and BOSS, respectively. Column (5): total number of young stars with RV measurements. Column (6): average RV of young stars listed in Column (5). Columns (7)--(8): same as Columns (5)--(6), but restricted to young stars with RV uncertainties of $\sigma_{\rm RV}<5$ \kms. Columns (9)--(10): same as Columns (5)--(6), but restricted to young stars with $\sigma_{\rm RV}<3$ \kms. Column (11): average RV of molecular gas traced by \lco, identical to Column (11) of Table~\ref{tab:basic_para}. All RVs in this table are given in the LSR frame.
}
\end{deluxetable*}

\section{Testing Different Reference Centers}\label{app:test_AC}

\begin{figure}[ht!]
	\centering
	\figurenum{C1}
	\includegraphics[width=0.95\textwidth]{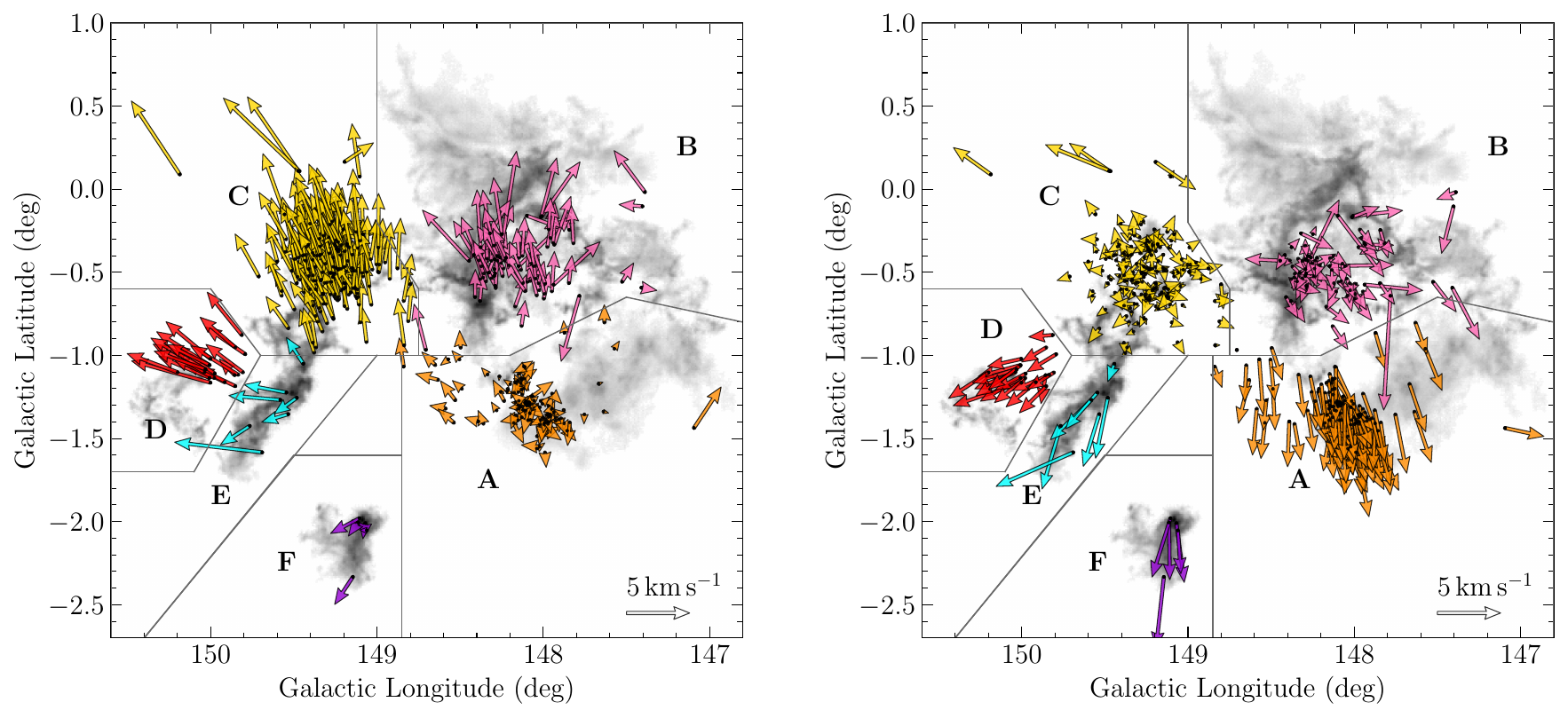}
	\caption{Residual tangential motions of young stars relative to the mean motion of the young star clusters in subregions A (left) and C (right). Symbols and colors are the same as in Figure~\ref{fig:2dmotion} (b). \label{fig:appC1}}
\end{figure}

\begin{figure*}[t!]
	\centering
	\figurenum{C2}
	\includegraphics[width=1\textwidth]{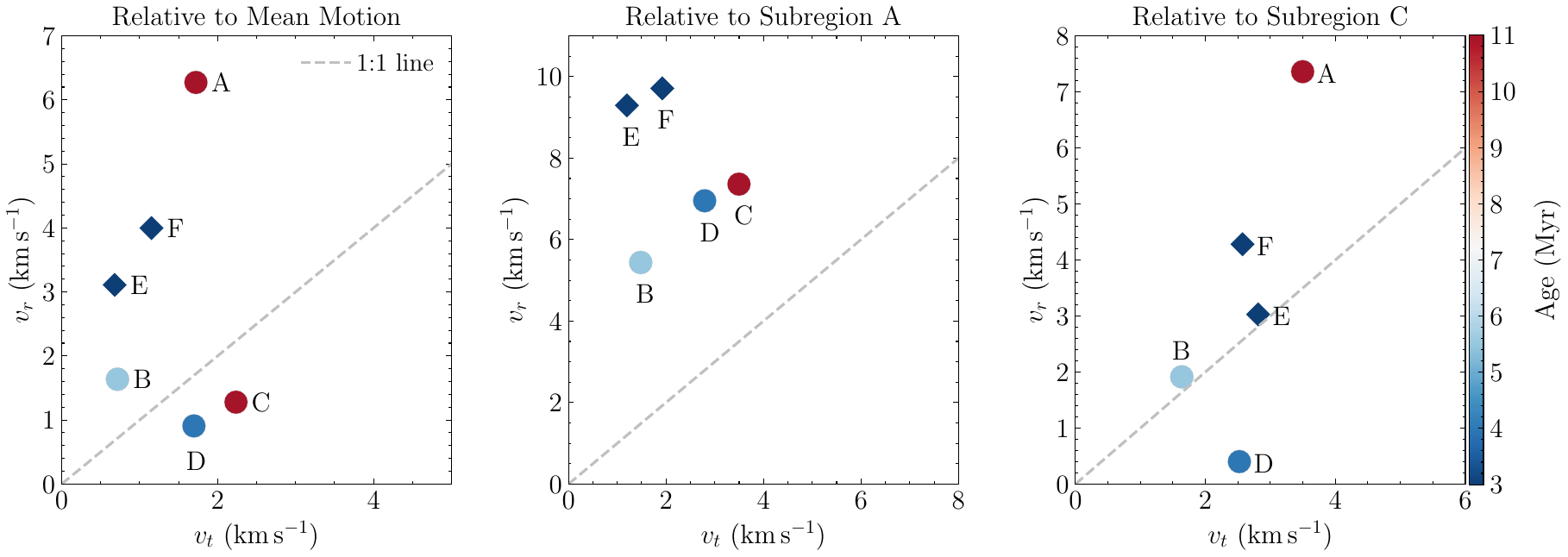}
	\caption{Radial ($v_r$) and tangential ($v_t$) components of {the} 3D relative motions of the subregions. From left to right, the reference center is defined by the mean motion of all six subregions, subregion A, and subregion C. Symbols and colors are as in Figure~\ref{fig:3d_project}. Positive $v_r$ values indicate radial outward motion. The dashed line marks $v_r=v_t$. See Section~\ref{sec:mor_motion}. \label{fig:vr_vt}}
\end{figure*}

To test the impact of the reference-center choice, we repeat the analysis using subregions A and C as alternative centers. These subregions host the two oldest stellar clusters in our sample and may therefore have served as sources of stellar feedback that influenced the younger subregions (see Section~\ref{sec:ages}). Figure~\ref{fig:appC1} shows the resulting residual tangential velocities of young stars, which still present a coherent expanding pattern. We further compare the radial ($v_r$) and tangential ($v_t$) components of the 3D relative velocities of the subregions (Figure~\ref{fig:vr_vt}). In all cases considered, the subregions show positive $v_r$ relative to the adopted center, suggesting that the global expansion of S205 is insensitive to the choice of reference center.

\clearpage

\bibliography{mybib}{}
\bibliographystyle{aasjournal}

\end{document}